\documentclass[aps,showpacs,twocolumn]{revtex4}
\usepackage{graphicx}

\begin{document}

\newcommand{\lw}[1]{\smash{\lower2.ex\hbox{#1}}}


\title{
Spacetime structure of static solutions
in Gauss-Bonnet gravity: charged case
} 

\author{
Takashi Torii$^{(1)}$ and Hideki Maeda$^{(2)}$
}

\address{ 
$^{(1)}$Graduate School of Science,
Waseda University, Shinjuku-ku, Tokyo 169-8555, Japan
}
\address{
$^{(2)}$Advanced Research Institute for Science and Engineering,
Waseda University, Shinjuku-ku, Tokyo 169-8555, Japan
}

\date{\today}

\begin{abstract}
We have studied spacetime structures of static solutions in
the $n$-dimensional Einstein-Gauss-Bonnet-Maxwell-$\Lambda$ system.
Especially we focus on effects of the Maxwell charge.
We assume that the Gauss-Bonnet coefficient $\alpha$ is non-negative and
$4{\tilde \alpha}/\ell^2\leq 1$ in order to define the relevant vacuum state.
Solutions  have the $(n-2)$-dimensional Euclidean sub-manifold whose  curvature  is $k=1,~0$, or $-1$. 
In Gauss-Bonnet gravity, solutions are classified into plus and minus branches. 
In the plus branch all solutions have the same  asymptotic structure as those in general relativity with a negative cosmological constant. 
The charge affects  a  central region of the spacetime.
A branch singularity appears at the finite radius $r=r_b>0$ for any mass parameter. 
There the Kretschmann invariant behaves as 
$O((r-r_b)^{-3})$, which is much milder than
divergent behavior of the central singularity in general relativity $O(r^{-4(n-2)})$.
In the $k=1$ and $0$ cases plus-branch solutions  have no horizon.
In the $k=-1$ case, the radius of a horizon is restricted as 
$r_h<\sqrt{2\tilde{\alpha}}$ ($r_h>\sqrt{2\tilde{\alpha}}$) in the plus (minus)
branch. 
Some charged black hole solutions  have no inner horizon in Gauss-Bonnet gravity. 
There are topological black hole solutions with zero and negative mass in the plus branch regardless of the sign of the cosmological constant. 
Although there is a maximum mass for black hole solutions  in the plus branch for $k=-1$ in the neutral case, no such maximum exists in the charged case.
The solutions in the plus branch with $k=-1$ and  $n\geq6$ have  an ``inner" black hole, and inner and the ``outer" black hole horizons.
In the  $4{\tilde \alpha}/\ell^2=1$ case,
only a positive mass solution is allowed, otherwise the metric function
takes a complex value.
Considering the evolution of black holes, we briefly discuss
a classical discontinuous transition from one black hole spacetime to another. 
\end{abstract}




\maketitle

\section{Introduction}
\label{Intro}

Recently a large number of studies in the field of black hole physics have focused
on the question of
higher dimensions. One of the reasons for this is the fascinating 
new picture of our Universe called the braneworld
universe~\cite{large,Randall,DGP}.
Because the fundamental scale could be about TeV
in this senario, these models suggest that creation of tiny black holes in
the upcoming linear hadron collider will be possible~\cite{Bhformation}. This would be  strong evidence of extra-dimensions.

Underlying the   braneworld model is an
underlying fundamental theory, superstring/M-theory~\cite{Lukas}, which is the most  promising candidate to produce a quantum theory of gravity. 
Although superstring/M-theory has been highly elaborated, it is not enough to understand black hole physics in the string context. Hence  at present
to take string effects perturbatively into classical gravity is one 
approach to the study of quantum gravity effects.  
With the restriction that the tension of a string be  large as compared to the energy scale of other variables, i.e., in the $\alpha'$-expansion, where $\alpha'$
is inverse string tension, the  Gauss-Bonnet terms appear as the first curvature correction term to general relativity~\cite{Gross}.
These are ghost-free combinations~\cite{Zwie} and have quasi-linear properties~\cite{GB_brane_linear}. 

Against this background we studied   spacetime structures of static solutions  in
the $n$-dimensional Einstein-Gauss-Bonnet-$\Lambda$ system systematically~\cite{TM}. 
The black hole solutions in Gauss-Bonnet gravity were first discovered by Boulware
and Deser~\cite{GB_BH} and Wheeler~\cite{Wheeler_1}, independently. 
Since then many types of black solutions have been  intensively studied~\cite{Torii,TM}.
These solutions have a good many characteristic properties which cannot be seen for black hole solutions  in general relativity.
Let us briefly review the characteristic properties of the uncharged solutions in 
Gauss-Bonnet gravity.
We assume that the Gauss-Bonnet coefficient $\alpha$, which corresponds to $\alpha'$, is non-negative, and a cosmological constant is either positive, zero, or negative.
The general solutions are classified into plus and minus branches. 
For the negative mass parameter, a new type of singularity called  a branch singularity appears at the non-zero finite 
radius $r=r_b>0$. The divergent behavior around the singularity in Gauss-Bonnet gravity is milder than that around the central singularity in general relativity.
A black hole solution with   zero or negative mass  exists in the
plus branch even for a zero or positive cosmological constant.
There is also an extreme black hole solution with positive mass, in spite of the
lack of an electromagnetic charge.

In this paper we extend the previous investigations by including the Maxwell charge.
In general relativity, the Maxwell charge affects the structure of the 
singularity, the 
number of horizons, and thermodynamical properties. 
In general relativity, if we add the charge to a Schwarzschild black hole, an inner horizon appears, and the singularity becomes timelike so that the spacetime represents  the 
Reissner-Nordstr\"om (RN) black hole. 
Since the singularity is timelike, a test observer can see it. This example seems to 
counter the   cosmic censorship hypothesis (CCH), proposed by Penrose~\cite{penrose1969,penrose1979}.
The CCH  asserts that the
singularities formed in generic gravitational collapse of physical
matter should be covered by the event horizon and remain invisible. There are two
versions of the CCH. One of them is  the weak cosmic censorship hypothesis,
which asserts that observers at infinity
should not see singularities. The other is the strong one, which asserts
that any observers
should not see them, i.e., 
the spacetime is globally
hyperbolic.
Nevertheless, the RN solution
should not be considered as a counterexample to the strong version of the CCH. Penrose
demonstrated that perturbations originating outside the black hole would
be blue-shifted infinitely at the inner horizon and that the inner horizon suffers from blue-sheet
instability~\cite{penrose1968}.  After a large number of analyses,
 it has been clarified that this type of inner horizon is unstable against the perturbations and
transforms into a null, weak scalar curvature singularity~\cite{pi1990,brady1999}. 
The general proof of the CCH is, however, far from complete, and many
counterexample candidates  have been proposed in the framework of
general relativity~\cite{Harada}. 
Then the following question arises: Can the string corrections 
support a system sufficiently for  the CCH to hold?
Furthermore, the charge may play an important roll in the braneworld and adS/CFT correspondence~\cite{adS/CFT}. In fact, in the braneworld model in general relativity, a cosmological constant may be self-tuned to solve the cosmological constant problem in
charged bulk spacetime~\cite{Caki}. 
From these points of view, investigations of the charged solution are important.

In Sec.~\ref{Model}, we introduce our model and show solutions  which are
generalizations of the originals of Boulware and Deser's. In Sec.~\ref{EML}, we
review the charged solutions in general relativity for comparison. 
In Sec.~\ref{EGBML}, the general properties of the solutions in the Einstein-Gauss-Bonnet-Maxwell-$\Lambda$
system with $4\tilde{\alpha}/\ell^2<1$, whose meaning is given in the text, are investigated. 
In Sec.~\ref{EGBM}, we show $\tilde{M}$-$r$ diagrams and study the number of horizons in each solution.
The global structures of all
solutions are summarized in tables.
Sec.~\ref{EGBML2} is devoted to the analysis of the  special case where
$4\tilde{\alpha}/\ell^2=1$.
In Sec.~\ref{Conclusion}, we present conclusions and discuss related issues.
Throughout this paper we use units such that $c=\hbar=k_B=1$. As for notation and
conversion we follow Ref.~\cite{Gravitation}. 
The Greek indices run $\mu$, $\nu=0,1, \cdots, n-1$.

\section{Model and solutions}
\label{Model}

We start with the following $n$-dimensional ($n\geq 4$) action
\begin{equation}
\label{action}
S=\int d^nx\sqrt{-g}\biggl[ 
\frac{1}{2\kappa_n^2}
(R-2\Lambda+\alpha{L}_{GB}
) \biggr]
+S_{\rm matter},
\end{equation}
where
$R$ and $\Lambda$ are the $n$-dimensional 
Ricci scalar and
the cosmological constant, respectively. $\kappa_n:=\sqrt{8\pi G_n}$, where $G_n$ is the $n$-dimensional gravitational constant. The Gauss-Bonnet Lagrangean ${L}_{GB}$ is
the combination of the Ricci scalar, the Ricci tensor $R_{\mu\nu}$, 
and the Riemann tensor $R_{\mu\nu\rho\sigma}$ as 
\begin{equation}
{L}_{GB}=R^2-4R_{\mu\nu}R^{\mu\nu}
+R_{\mu\nu\rho\sigma}R^{\mu\nu\rho\sigma}.
\end{equation}
In the four-dimensional case, the Gauss-Bonnet terms do not appear in the
equation of motion but contribute merely as surface terms. 
$\alpha$ is the coupling constant of the Gauss-Bonnet terms. This type of
action is derived from superstring theory in the low-energy limit~\cite{Gross}.
In that case 
$\alpha$ is regarded as the inverse string tension and positive definite.
For the stability of Minkowski
spacetime, we consider only the
case with $\alpha\geq 0$ in this paper.
We consider the Maxwell gauge field as a matter field, whose
action is
\begin{equation}
S_{\rm matter}=
-\frac{1}{4\pi g_c^2}\int d^nx\sqrt{-g}{F}_{\mu\nu}{F}^{\mu\nu},
\end{equation}
where $g_c$ is the Maxwell coupling constant.

The gravitational equation of the action (\ref{action}) is
\begin{equation}
{G}_{\mu\nu} +\alpha {H}_{\mu\nu} +\Lambda g_{\mu\nu}
= \kappa_n^2 
{T}_{\mu\nu},
\end{equation}
where 
\begin{eqnarray}
{G}_{\mu\nu}&:=&R_{\mu\nu}-{1\over 2}g_{\mu\nu}R,\\
{H}_{\mu\nu}&:=&2\Bigl[RR_{\mu\nu}-2R_{\mu\alpha}R^\alpha_{~\nu}
-2R^{\alpha\beta}R_{\mu\alpha\nu\beta}
\nonumber
\\
&& ~~~~
 +R_{\mu}^{~\alpha\beta\gamma}R_{\nu\alpha\beta\gamma}\Bigr]
-{1\over 2}g_{\mu\nu}{L}_{GB}.
\end{eqnarray}
The energy-momentum tensor of the Maxwell field is
\begin{eqnarray}
{T}_{\mu\nu}=\frac{1}{4\pi g_c^2}
\biggl({F}_{\mu\alpha}{F}_{\nu}^{~\alpha}
-\frac{1}{4}{g}_{\mu\nu}{F}_{\alpha\beta}{F}^{\alpha\beta}
\biggr).
\end{eqnarray}
The Maxwell equation is obtained as
\begin{equation}
\nabla_{\alpha}{F}^{\alpha\mu}=0.
\end{equation}

We assume static spacetime
and adopt the following line element:
\begin{equation}
\label{metric}
ds^2=-f(r)e^{-2\delta(r)}dt^2
+f^{-1}(r)dr^2+r^2d\Omega_{n-2}^2,
\end{equation}  
where $d\Omega_{n-2}^2=\gamma_{ij}dx^i dx^j$ 
is the metric of the $(n-2)$-dimensional Einstein space. 


The $(t, t)$ 
and $(r, r)$ components of the Einstein equation
provide $\delta'\equiv 0$, where a prime denotes the derivative with respect to $r$. By rescaling the time coordinate, 
we can always set
\begin{equation}
\label{delta}
\delta \equiv 0,
\end{equation}  
without loss of generality.

The Maxwell equation is written by using the electric field ${E}(r)$
defined by $E(r):=F_{rt}$ as
\begin{equation}
\frac{d}{dr} 
\bigl(r^{n-2}{E}\bigr)=0.
\end{equation}  
Here we do not consider the magnetic field. This equation is integrated to give 
\begin{equation}
{E}=\frac{Q}{r^{n-2}},
\end{equation}  
where the integration constant $Q$ is the electric charge.

The equation of the  metric function $f$ is then written as  
\begin{eqnarray}
\label{f-equation}
&&rf'-(n-3)(k-f)-\frac{n-1}{\ell^2}r^{2}
\nonumber \\
&& ~~~~~~~~~~~
+\frac{\tilde{\alpha}}{r^2}(k-f)
\Bigl[2rf'-(n-5)(k-f)\Bigr]
\nonumber \\
&& ~~~~~~~
=-\frac{\kappa_n^2Q^2}{4(n-2)\pi r^{2n-6}},
\end{eqnarray}  
where  
$\tilde{\alpha}:=(n-3)(n-4)\alpha$ and $\Lambda=-(n-1)(n-2)/(2\ell^2)$.
Note that a negative (positive) $\Lambda$ gives a positive (negative) $\ell^2$. 
$k$ is the curvature
of the $(n-2)$-dimensional Einstein space and takes $1$ (positive curvature), 
$0$ (flat),
or $-1$ (negative curvature).
The general solution of Eq.~(\ref{f-equation}) is obtained as
\begin{equation}
\label{f-eq}
f
=k+\frac{r^2}{2\tilde{\alpha}}
\Biggl\{1
\mp\sqrt{1+4\tilde{\alpha}\biggl[
\frac{\tilde{M}}{r^{n-1}}
-\frac{1}{\ell^2}-\frac{(n-3)\tilde{Q}^2}{2r^{2(n-2)}}
\biggr]}\Biggr\},
\end{equation}  
where
\begin{eqnarray}
\tilde{M}&:=&\frac{16\pi G_n M}{(n-2)\Sigma_{n-2}^k},\\
\tilde{Q}^2&:=&\frac{\kappa_n^2 Q^2}{2(n-2)\pi g_c^2}.
\end{eqnarray}  
The integration constant $M$ is proportional to the mass of a 
black hole for the black hole spacetime. Although
we will also consider solutions which do not represent black holes, we call $M$
the mass of the solution.
$\Sigma_{n-2}^k$ is the volume of the unit $(n-2)$-dimensional Einstein
space. 
There are two families of solutions which correspond to the sign in front of the
square root in Eq.~(\ref{f-eq}). We call the family which has a minus (plus) sign
the minus- (plus-) branch solution.
By introducing new variables as $r:=r/\ell$, 
$\bar{M}:=\tilde{M}/\ell^{n-3}$,
$\bar{Q}:=\tilde{Q}/\ell^{n-3}$, and
$\bar{\alpha}:=\tilde{\alpha}/\ell^2$,
the curvature radius $\ell$ is scaled out when the cosmological constant is
non-zero. 
In the $\Lambda=0$ case the variables are scaled as
$r:=r/\sqrt{\tilde{\alpha}}$, 
$\bar{M}:=\tilde{M}/\sqrt{\tilde{\alpha}}$, and
$\bar{Q}:=\tilde{Q}/\sqrt{\tilde{\alpha}}$.

The global structure of the spacetime is characterized by properties of the 
singularities, horizons, and infinities. 
A horizon is a null hypersurface defined by $r=r_h$ such that $f(r_h)=0$
with finite curvatures, where $r_h$
is a constant horizon radius.  
In this paper, we call a horizon on which $df/dr|_{r=r_h}>0$ a black hole horizon~\cite{Penrose}.
If there is a horizon inside of a black hole horizon, and if it satisfies 
$df/dr|_{r=r_h}<0$, we call it an inner horizon.
If a horizon satisfies $df/dr|_{r=r_h}<0$, and if it is the outermost
horizon, we call it a cosmological horizon.  We call a horizon on which
$df/dr|_{r=r_h}=0$ a degenerate horizon.  
Among them, if the first non-zero derivative coefficient is positive, i.e.,
$d^pf/dr^p|_{r=r_h}>0$ and $d^qf/dr^q|_{r=r_h}=0$ for any $q<p$, $q, p \in {\rm N}$,
we call it a degenerate black hole horizon.
A solution with the degenerate horizon is called extreme. 
The solutions in this paper are classified into two types by horizon existence. The first is a black hole solution which has
a black hole horizon.  
A solution which does not have a black hole horizon but has a locally naked singularity is a globally naked solution.

Since there are many parameters in our solution, such as $\tilde{\alpha}$,
$\ell^2$  (or $\Lambda$), $\tilde{M}$, $\tilde{Q}$, $k$, and $\pm$ branches, 
the analysis should be
performed systematically. In this paper we employ $\tilde{M}$-$r$ diagrams, which
are explained in Ref.~\cite{TM}.

\section{Static solutions in the Einstein-Maxwell-$\Lambda$ system}
\label{EML}

In Gauss-Bonnet gravity, two branches of the static solution appear. The minus-branch solution reduces to the solution in the Einstein-Maxwell-$\Lambda$ system in the  limit of $\alpha\to 0$. The metric function $f$ becomes 
\begin{eqnarray}
\label{einstein}
f=k-\frac{\tilde{M}}{r^{n-3}}
+\frac{(n-3)}{2}\frac{\tilde{Q}^2}{r^{2n-6}}
+\frac{r^2}{\ell^2}.
\end{eqnarray}  
On the other hand, $f$ diverges for the plus-branch solution in this limit, and
there is no counterpart in the Einstein-Maxwell-$\Lambda$ system.
Before proceeding to the case with the Gauss-Bonnet terms, we first
summarize the spacetime structure of the static
solutions described by Eqs.~(\ref{metric}), 
(\ref{delta}), and (\ref{einstein}) in the Einstein-Maxwell-$\Lambda$ system for comparison.
Being interested in the charged solution,
we assume $\tilde{Q}\ne 0$ throughout this paper.

\begin{figure}[tbp]
\includegraphics[width=.50\linewidth]{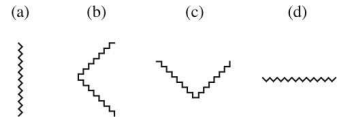}
\caption{
The conformal diagrams around  a singularity. Wavy lines denote  singularities.
There are time-reversed diagrams for (c) and (d).
}
\label{penrose-center}
\end{figure}

There is a singularity  at the center ($r=0$). 
The Kretschmann invariant behaves
\begin{eqnarray}
{\cal I}&:=&R_{\mu\nu\rho\sigma}R^{\mu\nu\rho\sigma}
\nonumber
\\
&=& O\Bigl(\frac{\tilde{Q}^2}{r^{4n-8}}\Bigr),
\label{kretchemann_e}
\end{eqnarray}  
around the center. 
The metric function  $f$ is positive around the center by Eq.~(\ref{einstein}), and 
the tortoise coordinate defined by 
\begin{equation}
r^{\ast}:=\int^{r} f^{-1}dr,
\label{tortoise}
\end{equation}  
is finite at the center. 
Hence the  singularity is always timelike (Fig.~\ref{penrose-center}(a)).

\begin{figure}[tbp]
\includegraphics[width=.52\linewidth]{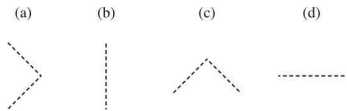}
\caption{
The conformal diagrams of the region of $r \to \infty$.
There are time-reversed diagrams for (c) and (d).
}
\label{penrose-infty}
\end{figure}

The structure of the infinity depends on the dominant term of Eq.~(\ref{einstein}) for 
$r \to \infty$. 
When there is a non-zero cosmological constant, the dominant term is $r^2/\ell^2$. 
When $\Lambda>0$ ($\Lambda<0$), the spacetime asymptotically
approaches  the de Sitter (dS) (anti-de Sitter (adS)) spacetime, whose region is expressed by the
conformal diagram Fig.~\ref{penrose-infty}(d)
(Fig.~\ref{penrose-infty}(b)). The second leading term is the
curvature term $k$. 
If $\Lambda=0$ and $k=1$, the spacetime is asymptotically flat, whose region is expressed by the conformal diagram Fig.~\ref{penrose-infty}(a).
If $\Lambda=0$ and $k=-1$, the conformal diagram of the infinity is Fig.~\ref{penrose-infty}(c).
When $\Lambda=k=0$, the $\tilde{M}$ term is dominant.
For $\tilde{M}>0$ ($\tilde{M}<0$), the infinity is expressed by Fig.~\ref{penrose-infty}(c) (Fig.~\ref{penrose-infty}(a)).
When $\Lambda=k=\tilde{M}=0$, the $\tilde{Q}^2$ term is dominant, and
 the infinity is expressed by Fig.~\ref{penrose-infty}(a).

The horizon plays an important role in determining the spacetime structure.
The existence of the horizon is clearly visualized on the $\tilde{M}$-$r$ diagram,
especially in the case where there are many parameters.
By Eq.~(\ref{einstein}), the $\tilde{M}$-$r_h$ relation in the 
Einstein-Maxwell-$\Lambda$ system is 
\begin{eqnarray}
\label{M-horizon-E}
\tilde{M}=r_h^{n-1}\biggl[\frac{1}{\ell^2}
+\frac{(n-3)}{2}\frac{\tilde{Q}^2}{r_h^{2n-4}}
+\frac{k}{r_h^2}
\biggr].
\end{eqnarray}  
We consider here that the values of $k$, $\ell$, and $\tilde{Q}$ are fixed so that this relation becomes a
curve on the $\tilde{M}$-$r$ diagram.

From the degeneracy condition of the horizon $f(r_{ex})=df/dr|_{r=r_{ex}}=0$, where $r_{ex}$ is the radius of the degenerate horizon, we can show that the $\tilde{M}$-$r_h$ curve on the diagram becomes vertical at this point. 
However, all the horizons at the vertical point are not necessarily degenerate. As  is
shown in Ref.~\cite{TM}, if 
\begin{eqnarray}
\label{vertical}
\frac{\partial \bar{f}}{\partial \tilde{M}} \bigg|_{r=r_h} =\infty,
\end{eqnarray}  
where $\bar{f}(r, \:\tilde{M}):= f(r)$,
the horizon at the vertical point can be nondegenerate.
Although this condition (\ref{vertical}) is satisfied only if $r_h \to 0$ limit in general relativity,
we will find below that there is no such solution. Hence the horizon at any vertical point in the Einstein-Maxwell-$\Lambda$ system
is a degenerate horizon.

The mass $\tilde{M}$ and the charge $\tilde{Q}^2$ of the
extreme solution are
\begin{eqnarray}
\label{M-extreme-E}
\tilde{M}&=&\tilde{M}_{ex}
=2r_{ex}^{n-1}\biggl[\frac{(n-2)}{(n-3)\ell^2}
+\frac{k}{r_{ex}^2}
\biggr],\\
\label{Q-extreme-E}
\tilde{Q}^2 
&=&\tilde{Q}_{ex}^2
=\frac{2r_{ex}^{2(n-2)}}{(n-3)^2}
\Bigl[
\frac{n-1}{\ell^2}
+\frac{(n-3)k}{r_{ex}^2}
\Bigr],
\end{eqnarray}  
respectively.
By Eq.~(\ref{Q-extreme-E}), there are  extreme solutions only in the case where
$\Lambda<0$ and/or $k=1$.
On the $\tilde{M}$-$r$ diagram the extreme solution is expressed by the point where
two horizons coincide. 
When
$d^2\tilde{M}/dr_h^{~2}=0$ is satisfied at the finite radius  $r_h=r_d$, the
three horizons degenerate, and we call the horizon a doubly degenerate horizon.

\begin{widetext}

\begin{figure}
\includegraphics[width=.70\linewidth]{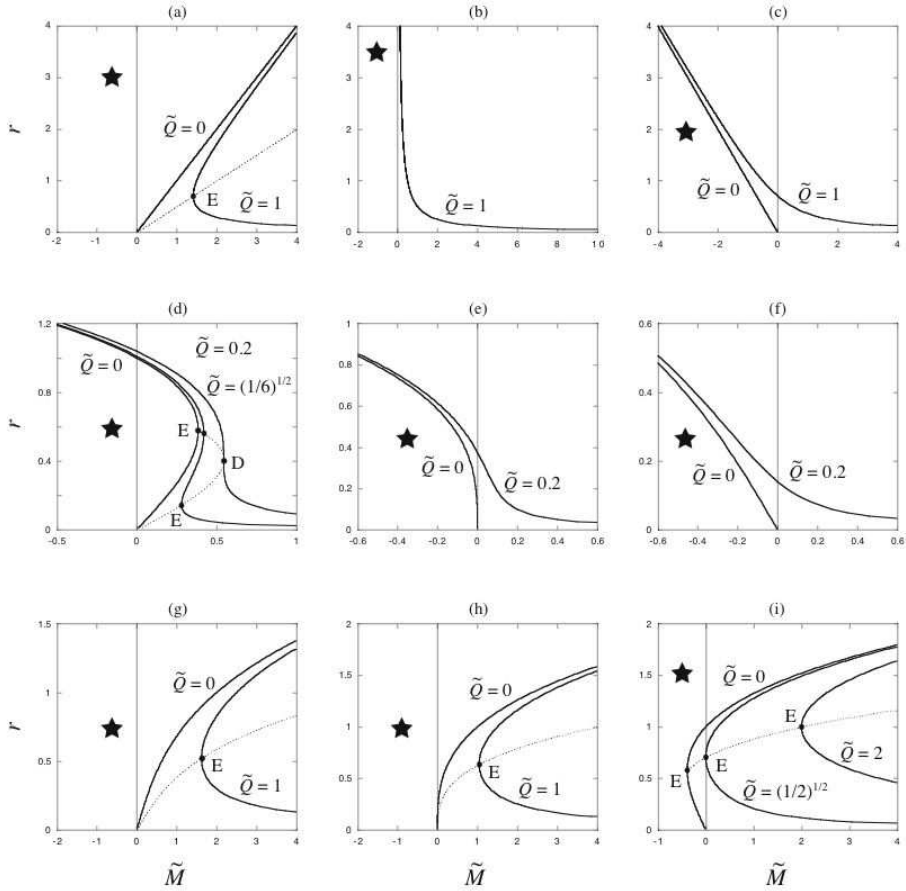}
\caption{
The $\tilde{M}$-$r$ diagrams for the static
solutions in the
four-dimensional Einstein-Maxwell-$\Lambda$ system. 
The diagrams in the upper, middle, and lower rows are the 
$1/\ell^2=0$ (zero cosmological constant), 
$1/\ell^2=-1$ (positive cosmological constant), and
$1/\ell^2=1$ (negative cosmological constant) cases,
respectively.
The diagrams in the left, the middle, and the right
columns are the $k=1$, $0$, and $-1$ cases, respectively.
The $\tilde{M}$-$r_h$ relations (thick curves) and
the $\tilde{M}$-$r_{ex}$ relations (dotted curves) are plotted.
The dots with characters ``E" and ``D" imply the degenerate horizon 
and the  doubly degenerate horizon, respectively.
The region with a star which is bounded by the $\tilde{M}$-$r_h$
curve represents an untrapped region.
It changes from an untrapped region to a trapped region or vice versa by crossing
the $\tilde{M}$-$r_h$ curve. In cases with more than four dimensions,  the $\tilde{M}$-$r$ diagrams are qualitatively the same.
}
\label{fig:E_m-rh}
\end{figure}

\end{widetext}

We will examine a number of horizons of the solutions.
The $\tilde{M}$-$r$ diagrams are shown in Fig.~\ref{fig:E_m-rh},
and the global structures  are summarized  
in Table~\ref{E-penrose-charged} and \ref{E-penrose-charged_2}.

\subsection{$\Lambda=0$ case}

By Eq.~(\ref{M-horizon-E}) we find
\begin{eqnarray}
\label{RN}
r_h^{n-3}=\frac{1}{2k}\Bigl[\tilde{M}\pm 
\sqrt{\tilde{M}^2-2(n-3)k\tilde{Q}^2}
\Bigr]
\end{eqnarray}  
for $k\ne 0$, and
\begin{eqnarray}
\label{RN2}
r_h^{n-3}=\frac{(n-3)\tilde{Q}^2}{2\tilde{M}}
\end{eqnarray}  
for $k=0$.


In the $k=1$ case, the solution is the $n$-dimensional
RN solution~\cite{note_1}.
The $\tilde{M}$-$r$ diagram is shown in Fig.~\ref{fig:E_m-rh}(a).
For $\tilde{M}<\tilde{M}_{ex}$, where
$\tilde{M}_{ex} =2r_{ex}^{n-3}=\sqrt{2(n-3)\tilde{Q}^2}$,
the solution has no horizon and represents the spacetime with a globally naked
singularity.
For  $\tilde{M}=\tilde{M}_{ex}$ the solution is the $n$-dimensional extreme
RN black hole solution.
The term of the square root in Eq.~(\ref{RN}) vanishes, and a degenerate
horizon locates at
$r_{ex}=\bigl[(n-3)\tilde{Q}^2/2\bigr]^{1/2(n-3)}$.
For $\tilde{M}>\tilde{M}_{ex}$, the solution has  black hole and
inner horizons, which have radii with the plus and the minus signs in
Eq.~(\ref{RN}), respectively,
and  represents  the RN black hole spacetime.


In the $k=0$ case, 
the $\tilde{M}$-$r_h$ diagram is shown in Fig.~\ref{fig:E_m-rh}(b). 
For $\tilde{M}\leq 0$ the solution has no horizon and represents the spacetime with a globally naked singularity.
For $\tilde{M}> 0$ the solution  has  
a cosmological horizon and represents the spacetime with a globally naked singularity.


In the $k=-1$ case, 
the $\tilde{M}$-$r_h$ diagram is shown in Fig.~\ref{fig:E_m-rh}(c). 
The solution has a cosmological horizon for any value of
$\tilde{M}$ and represents the spacetime with a globally naked singularity.

\subsection{$\Lambda>0$ case}

\label{EMLk1}

In the $k=1$ case, the solution is the $n$-dimensional
RN-dS solution and has three horizons at most.
The $\tilde{M}$-$r_h$ diagram is shown in Fig.~\ref{fig:E_m-rh}(d).

When the charge of the solution is not as large as $0<|\tilde{Q}|<|\tilde{Q}_{d}|$, there are two vertical points in the
$M$-$r_h$ diagram. Here $\tilde{Q}_{d}$ is
the charge of the solution with a doubly degenerate horizon
\begin{eqnarray}
&& \tilde{Q}_{d}^2=\frac{2}{(n-2)(n-3)}r_{d}^{2(n-3)}, \\
&& r_{d}^2=\frac{(n-3)^2|\ell^2|}{(n-1)(n-2)}.
\end{eqnarray}  
These vertical point E's in Fig.~\ref{fig:E_m-rh}(d) correspond to two degenerate horizons with 
radii $r_{ex}^{(1)}$ and $r_{ex}^{(2)}$
($r_{ex}^{(1)}<r_{ex}^{(2)}$) calculated by solving Eq.~(\ref{Q-extreme-E})
with respect to $r_{ex}$,
respectively. The former is realized by the degeneracy of the black hole
and the inner horizons, while the latter is realized by the degeneracy of the black hole
and the cosmological horizons. 
We denote the mass of these solutions by
$\tilde{M}_{ex}^{(1)}$ and $\tilde{M}_{ex}^{(2)}$
($\tilde{M}_{ex}^{(1)}< \tilde{M}_{ex}^{(2)}$), respectively.
For $\tilde{M}<\tilde{M}_{ex}^{(1)}$, the solution has  a cosmological
horizon and represents the spacetime with a globally naked singularity. 
For $\tilde{M}=\tilde{M}_{ex}^{(1)}$, the solution has a degenerate black hole
horizon and a cosmological horizon and represents the extreme RN-dS black hole spacetime. 
For $\tilde{M}_{ex}^{(1)}<\tilde{M}<\tilde{M}_{ex}^{(2)}$, 
the solution has inner,  black hole, and  cosmological horizons.
The spacetime is the RN-dS black hole spacetime.
For $\tilde{M}=\tilde{M}_{ex}^{(2)}$, 
the solution has inner  and  degenerate horizons.
For $\tilde{M}>\tilde{M}_{ex}^{(2)}$, the solution has a cosmological horizon
and represents the  spacetime with a globally naked singularity.

When $|\tilde{Q}|=|\tilde{Q}_{d}|$, there is a solution with a
doubly degenerate horizon whose mass parameter is
\begin{equation}
\tilde{M}=\tilde{M}_{d}=\frac{4}{n-1}r_{d}^{n-3}.
\end{equation}  
The doubly degenerate horizon locates at $r=r_{d}$. This solution represents 
the spacetime with a globally naked singularity.
The ratio of the charge to the mass is
\begin{equation}
\frac{|\tilde{Q}_d|}{\tilde{M}_d}=\sqrt{\frac{(n-1)^2}{8(n-2)(n-3)}}.
\end{equation}  
The ratio takes the largest value $|\tilde{Q}_d| / \tilde{M}_d=3/4$ for $n=4$.
For $\tilde{M}\ne \tilde{M}_{d}$, the solution has a cosmological 
horizon and represents the  spacetime with a globally naked singularity.

When $|\tilde{Q}|>|\tilde{Q}_{d}|$, any solution has  a cosmological 
horizon and represents the  spacetime with a globally naked singularity.


In the $k=0$ and the $k=-1$ cases, the $\tilde{M}$-$r$ diagrams are shown in Figs.~\ref{fig:E_m-rh}(e) and \ref{fig:E_m-rh}(f), respectively.
They have qualitatively similar structures.
The solutions have one cosmological horizon for any value of 
$\tilde{M}$ and represent the  spacetime with a globally naked singularity.

\subsection{$\Lambda<0$ case}

The system with a negative cosmological constant have been intensively studied in the
adS/CFT and braneworld contexts. Furthermore there are
topological black hole solutions~\cite{Brown,Cai3}.


In the $k=1$ case, the solution is the RN-adS solution. 
The $\tilde{M}$-$r$ diagram is shown in Fig.~\ref{fig:E_m-rh}(g).
The $\tilde{M}$-$r$ diagram in the $k=0$ case is shown in Fig.~\ref{fig:E_m-rh}(h).  We can see that they have similar structures.
There is one vertical point in the $\tilde{M}$-$r_h$
diagrams. It corresponds to the degenerate horizon, whose  solution has the mass and the charge calculated by Eqs.~(\ref{M-extreme-E}) and (\ref{Q-extreme-E}). 
The mass  of both extreme solutions is positive.
For $\tilde{M}<\tilde{M}_{ex}$, the solution has no horizon and represents the  spacetime with a globally naked singularity.
For $\tilde{M}=\tilde{M}_{ex}$, the solution has an degenerate black hole horizon and represents the extreme black hole spacetime.
For $\tilde{M}>\tilde{M}_{ex}$, the solutions have inner and  black hole horizons. The solution with $k=1$
represents the $n$-dimensional RN-adS black hole spacetime.


In the $k=-1$ case,
the $\tilde{M}$-$r$ diagram is shown in Fig.~\ref{fig:E_m-rh}(i).
The solutions are classified into three types.
For $\tilde{M}<\tilde{M}_{ex}$, the solution has  no horizons and represents the  spacetime with a globally naked singularity.
For $\tilde{M}=\tilde{M}_{ex}$, the solution has a degenerate black hole
horizon and represents the  extreme black hole spacetime.
For $\tilde{M}>\tilde{M}_{ex}$, the solution has inner and  black hole
horizons and represents the black hole spacetime.
Due to the first term in the r.h.s. of Eq.~(\ref{M-horizon-E}), 
a black hole solution with zero or negative mass can exist.
As is seen from Fig.~\ref{fig:E_m-rh}(i), the black hole solution with the lightest mass for 
a fixed charge is the extreme solution. When the charge is small, the extreme black hole solution 
has negative mass.
When $|\tilde{Q}|=|\tilde{Q}_{ex,0}|$,
where
\begin{eqnarray}
\label{negative}
\tilde{Q}_{ex,0}^2 :=
\frac{2}{(n-2)(n-3)}\biggl(\frac{n-3}{n-2}\ell^2\biggr)^{n-3}.
\end{eqnarray}  
the extreme black hole solution has zero mass, and its  horizon radius is
\begin{eqnarray}
r_{ex}=\sqrt{\frac{n-3}{n-2}}\ell.
\end{eqnarray}  
For the charge $|\tilde{Q}|>|\tilde{Q}_{ex,0}|$, all the black hole solutions have positive mass.

\section{General properties 
of solutions in the Einstein-Gauss-Bonnet-Maxwell-$\Lambda$ system
}
\label{EGBML}

We proceed to the cases with Gauss-Bonnet terms. 
First, let us discuss the general properties of the static solutions in this
system.


For the well-defined theory with the relevant 
vacuum state ($\tilde{M}=\tilde{Q}=0$), where the metric function is
\begin{eqnarray}
\label{metric-f1}
f=k+\frac{r^2}{2\tilde{\alpha}}
\biggl(1\mp\sqrt{1-\frac{4\tilde{\alpha}}{\ell^2}}\biggr),
\end{eqnarray}  
we assume
\begin{eqnarray}
\label{vacuum}
\frac{4\tilde{\alpha}}{\ell^2} \leq 1.
\end{eqnarray}  
For $\Lambda\geq 0$, any
value of $\tilde{\alpha} \;(\geq 0)$ satisfies this condition while,  $0\leq\tilde{\alpha}\leq 1/4$
in the negative cosmological constant case.
In this and the next sections, we study the $4\tilde{\alpha}/\ell^2<1$ case.
The special case of $4\tilde{\alpha}/\ell^2=1$ will be investigated in
Sec.~\ref{EGBML2}.


From Eq.~(\ref{metric-f1}), the square of the effective curvature radius is
defined by
\begin{eqnarray}
\ell_{\rm eff}^2
:=\frac{\ell^2}{2}
\biggl(1\pm\sqrt{1-\frac{4\tilde{\alpha}}{\ell^2}}\biggr).
\end{eqnarray}  
The sign of  $\ell_{\rm eff}^2$ coincides with $\ell^2$ in the minus branch, while  $\ell_{\rm eff}^2$ is 
always positive  in the plus branch.
Hence the spacetime approaches the adS spacetime
asymptotically for $r\to \infty$ in the plus branch for $k=1$, even when the pure cosmological  constant is positive. 
In Sec.~\ref{EML}, we showed the structures of  the infinity of the solutions,
which depend on the value of $\ell^2$, in general relativity.
By replacing $\ell^2$  in the discussion in general relativity by $\ell_{\rm eff}^2$
the structures of  the infinity in the 
Einstein-Gauss-Bonnet-Maxwell-$\Lambda$ system are obtained.


In general relativity there is a central singularity for any charged solution. In Gauss-Bonnet gravity the inside
of the square root of Eq.~(\ref{f-eq}) vanishes at the finite radius $r=r_b$, and the
other type of singularity, called the branch singularity, appears there. 
The $\tilde{M}$-$r_b$ relation becomes
\begin{equation}
\label{branch-sing}
\tilde{M}=\tilde{M}_b
:=-\Bigl(1-\frac{4\tilde{\alpha}}{\ell^2}\Bigr)
\frac{r_b^{n-1}}{4\tilde{\alpha}}
+\frac{(n-3)\tilde{Q}^2}{2r_b^{n-3}}.
\end{equation}  
This relation is  independent of $k$.
In the neutral case $\tilde{Q}=0$, the branch singularity appears only for the
solution with negative mass~\cite{TM}, while  all the solutions have branch singularity in the charged case.
Around the branch singularity the Kretschmann invariant behaves as
\begin{equation}
\label{divbranch}
{\cal I}\sim O\bigl[(r-r_b)^{-3}\bigr].
\end{equation}  
From the condition (\ref{vacuum}),
$r_b$ decreases monotonically as
$\tilde{M}$ increases and asymptotically approaches  zero in the limit of
$\tilde{M} \to \infty$.  
It is noted that the divergent behavior of the branch singularity is milder than that of the central singularity in general relativity (see Eqs.~(\ref{kretchemann_e}) and (\ref{divbranch})).

The metric function $f$ behaves around the
branch singularity as
\begin{eqnarray}
&&f(r) \approx \biggl(k+\frac{r_b^2}{2\tilde{\alpha}}\biggl)  \nonumber 
\\
&& ~~\mp
\frac{r_b^2}{2\tilde{\alpha}}\sqrt{\frac{n-1}{r_b}\biggl(1-\frac{4{\tilde
\alpha}}{\ell^2}\biggl)+\frac{2(n-3)^2{\tilde\alpha}{\tilde
Q}^2}{r_b^{2n-3}}} 
(r-r_b)^{1/2}. \nonumber 
\\
&&
\end{eqnarray}  
Hence when $k=1$ or $0$, the singularity is timelike (Fig.~\ref{penrose-center}(a)).
When $k=-1$, the singularities are timelike  
and spacelike (Fig.~\ref{penrose-center}(d)) for
$r_b>\sqrt{2\tilde{\alpha}}$ and
$r_b<\sqrt{2\tilde{\alpha}}$, respectively.
In the special case of $k=-1$ and $r_b=\sqrt{2\tilde{\alpha}}$, the
singularities are spacelike and timelike  for the minus and plus branches,
respectively.


On the horizon the metric function $f$ vanishes: $f(r_h)=0$, and
we find
\begin{equation}
\pm\Bigl(1+\frac{2\tilde{\alpha}k}{r_h^2}\Bigr)
=\sqrt{1+4\tilde{\alpha}
\biggl[\frac{\tilde{M}}{r_h^{n-1}}-\frac{1}{\ell^2}
-\frac{(n-3)\tilde{Q}^2}{2r_h^{2(n-2)}}\biggr]}
>0,
\end{equation}  
where the signs in the l.h.s. represents the minus/plus branches.
Hence
\begin{eqnarray}
\label{??}
&& r_h^2<-2\tilde{\alpha}k, ~~~ (\mbox{\rm for the plus branch}),
\\
&& r_h^2>-2\tilde{\alpha}k, ~~~ (\mbox{\rm for the minus branch}).
\end{eqnarray}  
By these condition it is concluded that there is no horizon for $k=1, 0$ in the plus
branch.  
When $k=-1$, the horizon radius is restricted as
$r_h<\sqrt{2\tilde{\alpha}}$ ($r_h>\sqrt{2\tilde{\alpha}}$) in the plus (minus) branch. 

The $\tilde{M}$-$r_h$ relation is
\begin{eqnarray}
\label{M-horizon}
\tilde{M}=r_h^{n-1}\biggl[\frac{1}{\ell^2}
+\frac{(n-3)}{2}\frac{\tilde{Q}^2}{r_h^{2n-4}}
+\frac{k}{r_h^2}\Bigl(1+\frac{\tilde{\alpha} k}{r_h^2}\Bigr)
\biggr].
\end{eqnarray}  
For $k=0$ this relation is same as that in general relativity (\ref{M-horizon-E}).
For $k=-1$  the solution with $r_h=\sqrt{2\tilde{\alpha}}$
has a branch singularity at $r=r_B:=\sqrt{2\tilde{\alpha}}$
where
\begin{equation}
\label{M-branch}
\tilde{M}=\tilde{M}_B
:=\frac{(2\alpha)^{\frac{n-3}{2}}}{2}
\biggl[-\Bigl(1-\frac{4\tilde{\alpha}}{\ell^2}\Bigr)
+\frac{(n-3)\tilde{Q}^2}{(2\tilde{\alpha})^{n-3}}\biggr].
\end{equation}  
This implies that the sequence of solutions
is divided by the branch singularity into the plus and the minus
branches in the $k=-1$ case. 
It is seen that the $\tilde{M}$-$r_h$ curve in the $\tilde{M}$-$r$ diagram 
terminates at the $\tilde{M}$-$r_b$ curve just at the point we call the branch point (Point B
in Figs.~\ref{fig:M-r-6-charged} and \ref{fig:M-r-5-charged}). 
For the mass parameter with $\tilde{M}\ne\tilde{M}_B$, the
horizon radius  is always larger than  $r_b$.
While $\tilde{M}_B$ is always negative in the neutral case,
$\tilde{M}_B$ can be zero in the charged case  by tuning the charge $\tilde{Q}$ as
\begin{equation}
\label{QB}
\tilde{Q}^2={\tilde{Q}_{B,0}}^2
:=\frac{1}{n-3}
\biggl(1-\frac{4\tilde{\alpha}}{\ell^2}\biggr)
(2\tilde{\alpha})^{n-3}.
\end{equation}  
For $|\tilde{Q}|<|\tilde{Q}_{B,0}|$ ($|\tilde{Q}|>|\tilde{Q}_{B,0}|$), $\tilde{M}_B$ is negative
(positive).


As we will see in the next section, in the $\tilde{M}$-$r$ diagram of the 
$\tilde{M}$-$r_h$ curve there are some vertical points.
In Gauss-Bonnet gravity, 
\begin{equation}
\frac{\partial \bar{f}}{\partial\tilde{M}}\biggr|_{r=r_h}
=\mp \frac{1}{r_h^{n-3}}
\biggl\{1+4\tilde{\alpha}
\biggl[\frac{\tilde{M}}{r_h^{n-1}}-\frac{1}{\ell^2} 
- \frac{(n-3)\tilde{Q}^2}{2r_h^{2(n-2)}}\biggr]\biggr\}^{-\frac12}
\end{equation}  
is finite, so that 
the horizons at the vertical points are degenerate except for $r_h\to r_b$, where
the condition (\ref{vertical}) is satisfied.
However, the $\tilde{M}$-$r_h$ curve is tangent to the
$\tilde{M}$-$r_b$ curve at the branch point, and the $\tilde{M}$-$r_b$ curve monotonically decreases. Thus  the $\tilde{M}$-$r_h$ curve cannot be vertical at the branch point. As a result the horizons at all the vertical points are degenerate (for this analysis,  see Ref.~\cite{TM}.)

In the extreme cases, the relations between the  mass, the charge, and the horizon
radius  are written as
\begin{equation}
\label{M-extreme}
\tilde{M}=\tilde{M}_{ex}=2r_{ex}^{n-1}\biggl[\frac{(n-2)}{(n-3)\ell^2}
+\frac{k}{r_{ex}^2}
+\frac{(n-4)\tilde{\alpha}k^2}{(n-3)r_{ex}^4}
\biggr],
\end{equation}  
\begin{equation}
\label{Q-extreme}
\tilde{Q}^2 =\tilde{Q}_{ex}^2 =
\frac{2r_{ex}^{2(n-2)}}{(n-3)^2}
\biggl[
\frac{n-1}{\ell^2}
+\frac{(n-3)k}{r_{ex}^2}
+\frac{(n-5)\tilde{\alpha}k^2}{r_{ex}^4}
\biggr].
\end{equation}  
In five dimensions the last term in Eq.~(\ref{Q-extreme}) vanishes, and the charge
of the extreme solution has the same expression as that in general relativity. This implies that degenerate horizons appear
at the same radius as those appearing in general relativity if we
choose the  same cahrge  value. However, the mass of the extreme solution is
different from that in general relativity.
Eq.~(\ref{Q-extreme}) is rewritten as
\begin{eqnarray}
\label{ex-pot}
U(r_{ex})&:=&\frac{n-1}{\ell^2}r_{ex}^{2n-4}+(n-3)kr_{ex}^{2n-6}
\nonumber \\
&& \;\;\;
+\tilde{\alpha}k^2(n-5)r_{ex}^{2n-8}
\nonumber \\
&=&\frac{(n-3)^2}{2}\tilde{Q}^2.
\end{eqnarray}  
The function $U(r_{ex})$ is defined in the range $r_{ex}\geq 0$.
If $U(r_{ex})$ is negative semi-definite, there is no
extreme solution. If $U(r_{ex})$ is a monotonically
increasing function from zero to infinity, there is a
certain mass $\tilde{M}_{ex}$ where the solution becomes extreme
for any value of $\tilde{Q}$.
We denote the mass of the extreme solution in the plus (minus) branch as
$\tilde{M}_{ex}^{(+)}$ ($\tilde{M}_{ex}^{(-)}$). When more than one extreme solution exists in the same branch, we denote their mass as $\tilde{M}_{ex}^{(1+)}$,
$\tilde{M}_{ex}^{(2+)}$, $\cdots$ ($\tilde{M}_{ex}^{(1-)}$,
$\tilde{M}_{ex}^{(2-)}$, $\cdots$).

If $U(r_{ex})$ is not a monotonic function, there are extremal points, which
correspond to the solutions with a doubly degenerate horizon.
We should take care not to confuse these extremal points of the function $U(_{ex})$
with the  extremal solution with degenerate horizons.

Although there may be a horizon with more degeneracy in general,
in the present system up to doubly degenerate horizon can appear.
Hence examining the extremal points of $U(r_{ex})$, we find the condition and
location of the doubly degenerate horizon $r=r_d$ such that $dU/dr_{ex}|_{r_{ex}=r_d}=0$.
As the charge $\tilde{Q}$ is varied, the number of the 
roots $r_{ex}$ in Eq.~(\ref{ex-pot}) changes at the
extremal points of the function $U(r_{ex})$.
When $\Lambda=0$,
the doubly degenerate horizon appears at
\begin{eqnarray}
&&r_{d}^2=-\frac{(n-4)(n-5)}{(n-3)^2}k\tilde{\alpha}. 
\label{trila}
\end{eqnarray}  
Since the l.h.s. of this equation is positive,  $k$ must be $-1$ and $n\geq 6$.
Furthermore since $r_{d}<\sqrt{\tilde{\alpha}}$,
the doubly degenerate horizon exists only in the plus branch.
Then the mass and the charge of the solution is given by
Eqs.~(\ref{M-extreme}) and (\ref{ex-pot}) as
\begin{eqnarray}
&&
\tilde{M}_{d}=\frac{4}{n-5}r_{d}^{n-3},
\label{triM}
\\ 
&&
\tilde{Q}_{d}^2
:=\frac{2}{(n-3)^2}U(r_{d}) \nonumber
\\
&&
\;\;\;\;\;\;\;
=\frac{2}{(n-3)(n-4)}r_{d}^{2(n-3)}.
\label{triQ}
\end{eqnarray}  
When $\Lambda\ne 0$, the doubly degenerate horizon appears at
\begin{eqnarray}
\label{r-tri}
&&r_{d}^2=-\frac{(n-3)^2\ell^2B}{2(n-1)(n-2)},
\end{eqnarray}  
and the mass and  the charge of the solution become
\begin{eqnarray}
&&\tilde{M}_{d}=2r_{d}^{n-5}k^2
\biggl[\frac{4(n-4)\tilde{\alpha}}{(n-1)(n-3)}
-\frac{(n-3)^2\ell^2B}{(n-1)^2(n-2)k}
\biggr],
\label{M-tri}
\nonumber \\
&& \nonumber \\
&& \\
&& \nonumber \\
&&\tilde{Q}_{d}^2=r_{d}^{2(n-4)}k^2
\biggl[\frac{4(n-5)\tilde{\alpha}}{(n-2)(n-3)^2}
-\frac{(n-3)\ell^2B}{(n-1)(n-2)^2k}
\biggr],
\label{Q-tri}
\nonumber \\
&&
\end{eqnarray}  
where
\begin{eqnarray}
\label{BB}
B:=k\pm
|k|\sqrt{1-\frac{4\tilde{\alpha}(n-1)(n-2)(n-4)(n-5)}{(n-3)^4\ell^2}}.
\end{eqnarray}  
Note that the sign $\pm$ does not correspond to the signs of the solution branches
but is determined to make the r.h.s. of 
Eq.~(\ref{r-tri})  positive. If both signs give negative
$\ell^2 B$, there are two sets of parameters which give the solutions with a doubly degenerate horizon.
When $|\tilde{Q}_{d}|$ is single valued  (namely,
there is only one value of the charge which
gives a doubly degenerate horizon), two extreme solutions 
exist for $|\tilde{Q}|<|\tilde{Q}_{d}|$, while
no extreme solution exists for $|\tilde{Q}|>|\tilde{Q}_{d}|$.

\section{$M$-$r_h$ diagram and spacetime structure
}
\label{EGBM}

In the previous section, we discussed properties of the singularity,
 the infinity, and the horizons of the solutions in the
Einstein-Gauss-Bonnet-Maxwell-$\Lambda$
system. 
In this section we focus on the number of horizons and show the spacetime structures
by taking into account the above properties.
To clarify our considerations, the $\tilde{M}$-$r$
diagrams are shown in Figs.~\ref{fig:M-r-6-charged} ($n= 6$)
and \ref{fig:M-r-5-charged} ($n=5$).
The spacetime structures of the solutions are summarized in
Tables~\ref{E-penrose-charged_2}-\ref{GB-penrose-6-charged_2}.

\begin{widetext}

\begin{figure}
\includegraphics[width=.95\linewidth]{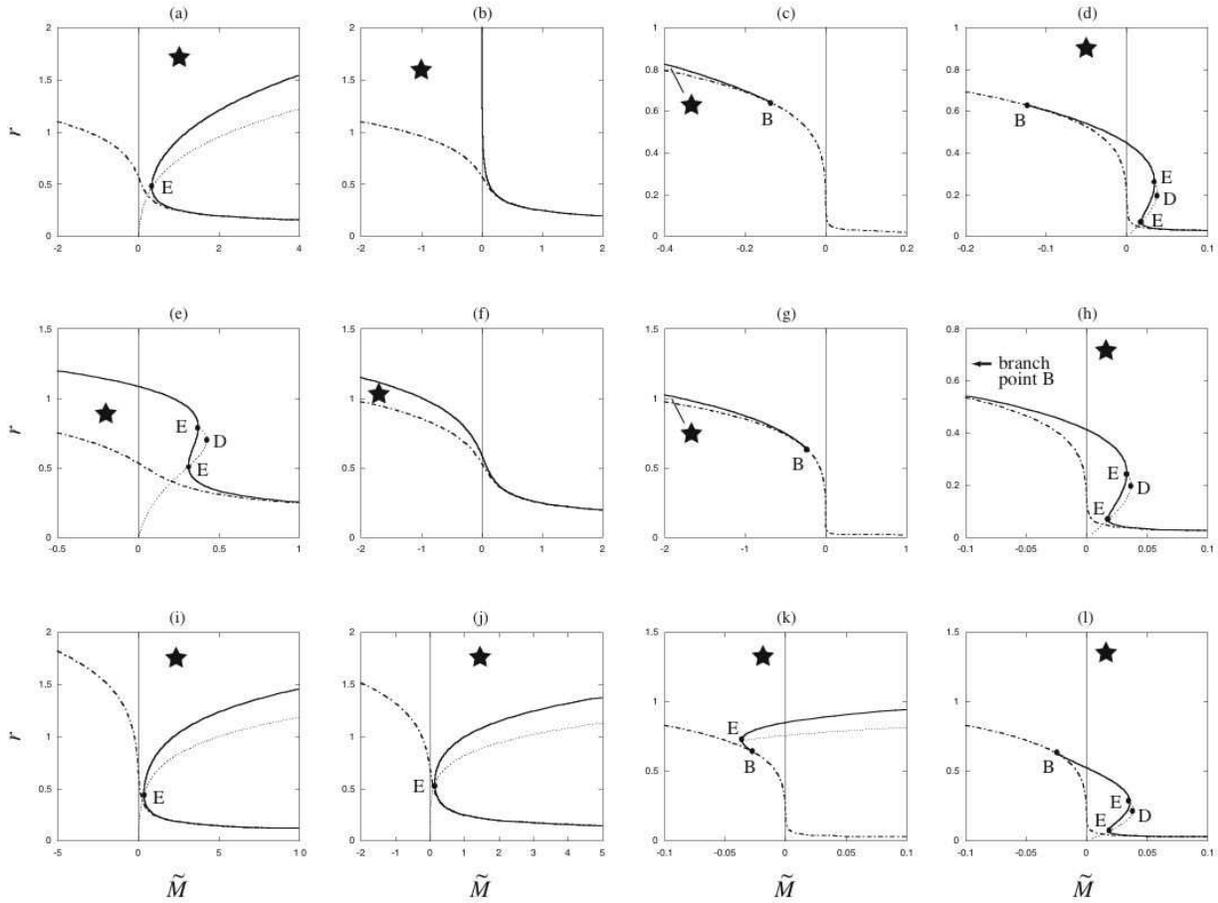}
\caption{
The $\tilde{M}$-$r$ diagrams of the static
solutions in the
six-dimensional Einstein-Gauss-Bonnet-Maxwell-$\Lambda$ system with $4\tilde{\alpha}/\ell^2\ne 1$. 
The diagrams in the upper, middle, and lower rows are the 
$1/\ell^2=0$ (zero cosmological constant), 
$1/\ell^2=-1$ (positive cosmological constant),  and
$1/\ell^2=1$ (negative cosmological constant) cases,
respectively.
The diagrams in the columns from the left-hand side
are the $k=1$ (the minus branch), $0$ (the minus branch), $-1$  (the
minus branch), and  $-1$  (the plus branch) cases, respectively. 
We show
the $\tilde{M}$-$r_h$ relations (thick solid curves) and
the $\tilde{M}$-$r_b$ relations (dot-dashed curves).  
We set $\tilde{\alpha}=0.2$ and $\tilde{Q}=0.1$ ($k=1, 0$)
and $\tilde{Q}=10^{-3}$ ($k=-1$).
We choose the value of the charge as $|\tilde{Q}|<|\tilde{Q}_{d}|$
in the $k=-1$ case. 
Otherwise, the two extreme solutions in the
plus branch coincide, and the $\tilde{M}$-$r_h$ relation becomes
single valued.
The dotted curve represents a sequence of the degenerate horizon ($\tilde{M}$-$r_{ex}$ relations)  by varying $\tilde{Q}$.
Below the $\tilde{M}$-$r_b$ line, there are no solutions.
The dots labeled ``E," ``D," and ``B" imply the degenerate horizon,
the doubly degenerate horizon, and the branch point, respectively.
See Fig.~\ref{fig:E_m-rh}  for the
meanings of the stars.
The $\tilde{M}$-$r$ diagrams of the higher-dimensional solutions
with $n\geq 6$ 
have similar configurations to these.
}
\label{fig:M-r-6-charged}
\end{figure}

\begin{figure}
\includegraphics[width=.90\linewidth]{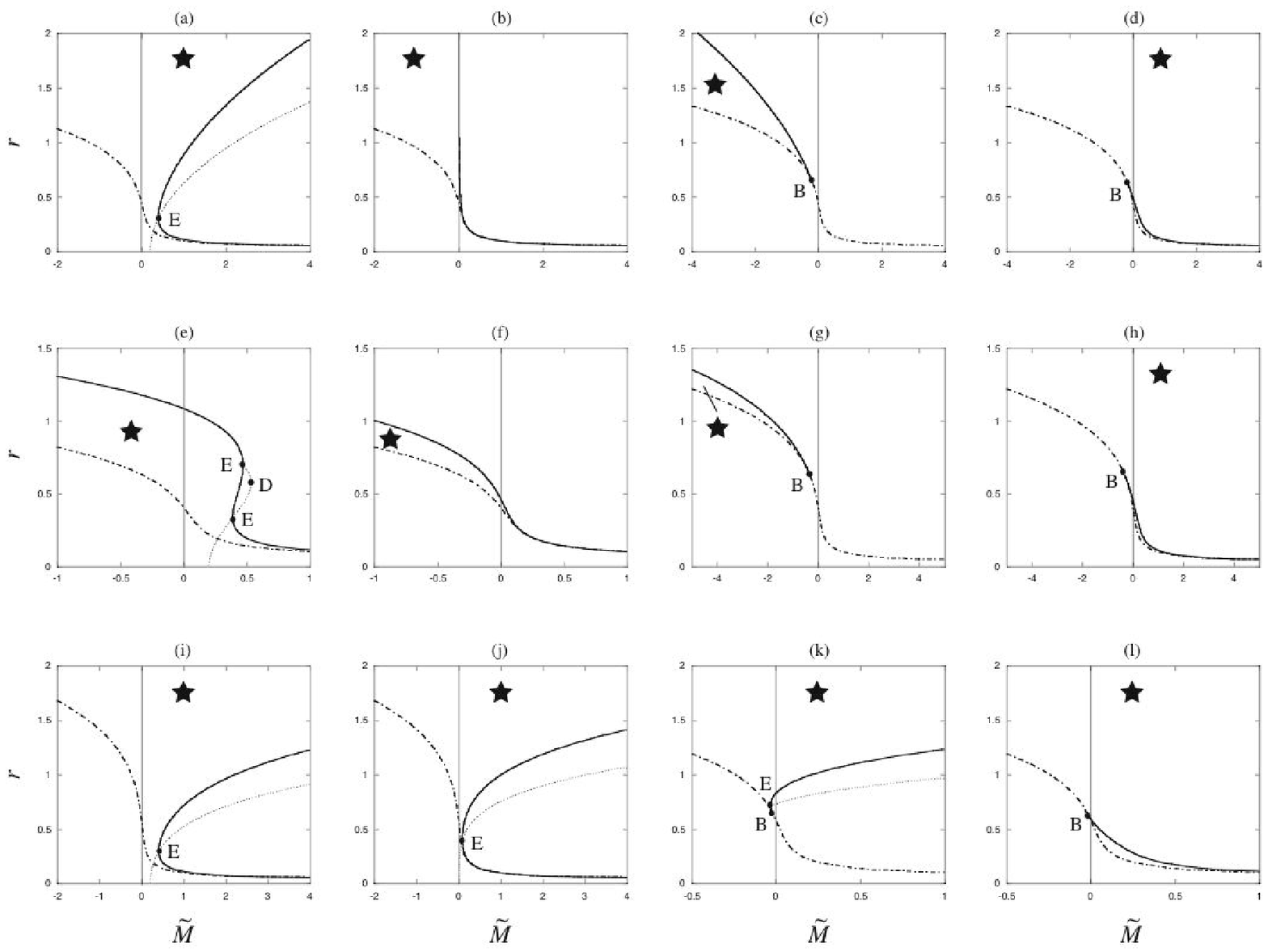}
\caption{
The $\tilde{M}$-$r$ diagrams of the static
solutions in the
five-dimensional Einstein-Gauss-Bonnet-Maxwell-$\Lambda$ system with $4\tilde{\alpha}/\ell^2\ne 1$. 
The diagrams in the upper, middle, and lower rows are the 
$1/\ell^2=0$ (zero cosmological constant), 
$1/\ell^2=-1$ (positive cosmological constant),  and
$1/\ell^2=1$ (negative cosmological constant) cases,
respectively.
The diagrams in the columns from the left-hand side
are the $k=1$ (the minus branch), $0$ (the minus branch), $-1$  (the
minus branch), and  $-1$  (the plus branch) cases, respectively. 
We show
the $\tilde{M}$-$r_h$ relations (thick lines) 
and the $\tilde{M}$-$r_b$ relations (dot-dashed lines).
We set $\tilde{\alpha}=0.2$ and $\tilde{Q}=0.1$.
Below the $\tilde{M}$-$r_b$ line, there are no solutions.
The dotted curve represents a sequence of the degenerate horizon ($\tilde{M}$-$r_{ex}$ relations)  by varying $\tilde{Q}$.
See Figs.~\ref{fig:E_m-rh} and \ref{fig:M-r-6-charged}  for the
meanings of the dots and the stars.
}
\label{fig:M-r-5-charged}
\end{figure}

\end{widetext}

\vspace{8mm}
\subsection{$\Lambda=0$}
\label{GB0}

\subsubsection{$k=1$  case}

The solution in the plus branch does not have
a  horizon and 
represents the spacetime with a globally naked singularity.

The $\tilde{M}$-$r$ diagrams of the minus-branch solutions are shown in 
Figs.~\ref{fig:M-r-6-charged}(a) ($n=6$)
and \ref{fig:M-r-5-charged}(a) ($n=5$).
Since the function $U(r_{ex})$ monotonically
increases from zero to infinity,
there is an extreme solution for any $\tilde{Q}$. The extreme solution has the positive mass $\tilde{M}_{ex}^{(-)}>0$. 
For $\tilde{M}<\tilde{M}^{(-)}_{ex}$, the solution has no horizon and represents the spacetime with a globally naked singularity.
For $\tilde{M}=\tilde{M}^{(-)}_{ex}$, the solution has a degenerate horizon and represents the extreme black hole spacetime. 
In the five-dimensional case, 
the mass and the horizon radius of the extreme black hole are
$\tilde{M}^{(-)}_{ex}=2|\tilde{Q}|+\tilde{\alpha}$ and 
$r_{ex}=\sqrt{|\tilde{Q}|}$, respectively.
For $\tilde{M}>\tilde{M}^{(-)}_{ex}$, the solution has inner  and  black hole horizons
and represents the black hole spacetime.

\subsubsection{$k=0$  case}

The solution in the plus branch does not have
a  horizon and 
represents the spacetime with a globally naked singularity.

The $\tilde{M}$-$r$ diagrams of the minus-branch solutions are shown in Figs.~\ref{fig:M-r-6-charged}(b) ($n=6$)
and \ref{fig:M-r-5-charged}(b) ($n=5$).
The $\tilde{M}$-$r_h$ relation becomes
$\tilde{M}=(n-3)\tilde{Q}^2/2r_h^{n-3}$, and the horizon radius diverges in the  $\tilde{M}\to 0+$ limit. 
For $\tilde{M}\leq 0$, the solution has no horizon and represents the spacetime with a globally naked singularity.
For $\tilde{M}>0$, the solution has a cosmological horizon and represents the spacetime with a globally naked singularity.

\subsubsection{$k=-1$  case}

In the $k=-1$  case, solutions in the plus branch can have horizons.
Since the function $U(r_{ex})$ has a positive maximum in the $n\geq6$ case,
there is a solution with a doubly degenerate horizon when
$|\tilde{Q}|=|\tilde{Q}_{d}|$, which is given by Eqs.~(\ref{trila}) and (\ref{triQ}).
This solution belongs to the plus branch since $r_d<\sqrt{2\tilde{\alpha}}$. When $|\tilde{Q}|<|\tilde{Q}_{d}|$, Eq.~(\ref{ex-pot}) has two roots, i.e., two types of degenerate horizons exist. Since the function $U(r_{ex})$ becomes positive for the range $0<r_{ex}<\sqrt{\tilde{\alpha}(n-5)/(n-3)}<\sqrt{2\tilde{\alpha}}$ in this case,
these extreme solutions also belong to the plus branch.
The mass of these extreme solutions is denoted as $\tilde{M}_{ex}^{(1+)}$ and $\tilde{M}_{ex}^{(2+)}$, where $\tilde{M}_{ex}^{(1+)} < \tilde{M}_{ex}^{(2+)}$, respectively.
On the other hand, the function $U(r_{ex})$ is negative semi-definite in $n=5$, and
there is no extreme solution.

Let us first consider the solutions in the minus branch.
The $\tilde{M}$-$r$ diagrams of the minus-branch solutions are shown in Figs.~\ref{fig:M-r-6-charged}(c) ($n=6$)
and \ref{fig:M-r-5-charged}(c) ($n=5$).
The diagrams in $n=5$ and higher-dimensional cases are qualitatively the same.
For $\tilde{M}<\tilde{M}_B$, the solution has a cosmological horizon and represents the spacetime with a globally naked singularity.
For $\tilde{M}\geq\tilde{M}_B$, the solution has no horizon,  and represents the spacetime with a globally naked singularity.
Here $\tilde{M}_B$ is positive (negative) when $|\tilde{Q}|>|\tilde{Q}_{B,0}|$
($|\tilde{Q}|<|\tilde{Q}_{B,0}|$), where $|\tilde{Q}_{B,0}|$ is given by Eq.~(\ref{QB}) as
\begin{equation}
\label{QE-5}
\tilde{Q}_{B,0}^2 =\frac{(2\tilde{\alpha})^{n-3}}{n-3}.
\end{equation}  

In the $n\geq6$ case, the $\tilde{M}$-$r$ diagram of the plus branch solutions is shown in Fig.~\ref{fig:M-r-6-charged}(d).
The configuration of the $\tilde{M}$-$r_h$ curve changes qualitatively depending on the value of the charge.
First we assume $|\tilde{Q}|<|\tilde{Q}_{d}|$. Then there are two  extreme solutions.
For $\tilde{M}\leq\tilde{M}_B$, the solution has no horizon and represents the spacetime with a globally naked singularity.
For $\tilde{M}_B<\tilde{M}<\tilde{M}_{ex}^{(1+)}$, the solution has a black hole horizon
and represents the black hole spacetime.
Since $|\tilde{Q}_d|<|\tilde{Q}_{B,0}|$ by Eqs.~(\ref{triQ}) and (\ref{QE-5}), $\tilde{M}_B$ is negative. 
So, the solution with $\tilde{M}_B<\tilde{M}<0$ represents the negative mass black hole.
For $\tilde{M}=\tilde{M}_{ex}^{(1+)}$, the solution has a degenerate and a black hole horizon and represents   the black hole spacetime.
For $\tilde{M}_{ex}^{(1+)}<\tilde{M}<\tilde{M}_{ex}^{(2+)}$, the solution has three horizons. Although the RN-dS black hole solution in general relativity also has  three horizons, the global  structure of the present solution differs from it.
By our definition the outermost and the innermost horizons of the three horizons are the black hole horizon, i.e., there are two black hole horizons, and the horizon between them is the inner horizon. For an observer locating in the untrapped region between the ``inner" black hole horizon, i.e., the innermost horizon,  and the inner horizon, the inner horizon would be seen as if it were the cosmological horizon.
For $\tilde{M}=\tilde{M}_{ex}^{(2+)}$, the solution has a black hole and a degenerate  black hole horizon and represents the extreme black hole spacetime. 
For $\tilde{M}>\tilde{M}_{ex}^{(2+)}$, the solution has a black hole  horizon and represents the black hole spacetime.

When $|\tilde{Q}|=|\tilde{Q}_{d}|$, there is a solution with a doubly degenerate horizon.
For $\tilde{M}\leq\tilde{M}_{B}$, the solution has no horizon and represents the   spacetime with a globally naked singularity.
For $\tilde{M}_{B}<\tilde{M}<\tilde{M}_{d}$, the solution has a black hole horizon and represents the black hole spacetime.
The solution with $\tilde{M}_B<\tilde{M}<0$ represents the negative mass black hole.
For $\tilde{M}=\tilde{M}_{d}$, where $\tilde{M}_{d}$ is given by Eqs.~(\ref{trila}) and (\ref{triM}), the solution has a doubly degenerate horizon. In general relativity, there is a solution with a doubly degenerate horizon as a special case of the RN-dS solution, as shown in Sec.~\ref{EMLk1}. That is not a black hole solution; while the present solution in Gauss-Bonnet gravity represents the black hole spacetime with a doubly degenerate horizon. 
The ratio of the charge to the mass becomes
\begin{eqnarray}
\frac{|\tilde{Q}_{d}|}{\tilde{M}_{d}}
=\sqrt{\frac{(n-5)^2}{8(n-3)(n-4)}}
<\frac{1}{\sqrt{2}}.
\end{eqnarray}  
For $\tilde{M}>\tilde{M}_{d}$, the solution has a black hole horizon and represents the black hole spacetime.

When $|\tilde{Q}|>|\tilde{Q}_{d}|$, no extreme solution exists.
For $\tilde{M}\leq\tilde{M}_{B}$,  the solution has no horizon  and represents the   spacetime with a globally naked singularity.
For $\tilde{M}>\tilde{M}_{B}$, the solution has a black hole horizon and represents the black hole spacetime.
If $|\tilde{Q}_{d}|<|\tilde{Q}|<|\tilde{Q}_{B,0}|$, the solution with $\tilde{M}_{B}<\tilde{M}<0$ represents the negative mass black hole.

In the $n=5$ case, the $\tilde{M}$-$r$ diagram of the plus branch solutions is shown in Fig.~\ref{fig:M-r-5-charged}(d). The qualitative behavior of the solutions in the minus branch is similar to that in the  $n\geq 6$ case, while that in the plus branch is a little different.
Since the function $U(r_{ex})$ is negative semi-definite in the $n=5$ case, there in no extreme solution. Hence, in the plus branch, 
for $\tilde{M}\leq\tilde{M}_{B}$, the solution has no horizon and represents the   spacetime with a globally naked singularity.
For $\tilde{M}>\tilde{M}_{B}$, the solution has a black hole horizon and represents the black hole spacetime.
If $|\tilde{Q}|<|\tilde{Q}_{B,0}|$, the solution with $\tilde{M}_{B}<\tilde{M}<0$ represents the  negative mass black hole.

\subsection{$\Lambda>0$}

\subsubsection{$k=1$  case}

The solution in the plus branch has no horizon and represents the spacetime with a globally
naked singularity. On the other hand all the solutions in the minus branch have horizons. The number of horizons depends on the charge and the mass.

The $\tilde{M}$-$r$ diagrams of the minus-branch solutions are shown in Figs.~\ref{fig:M-r-6-charged}(e) ($n=6$)
and \ref{fig:M-r-5-charged}(e) ($n=5$).
Since the function $U(r_{ex})$ has the positive maximum, there is a solution with a doubly degenerate horizon, whose mass and charge are given by Eqs.~(\ref{M-tri}) and (\ref{Q-tri}), respectively.
When $|\tilde{Q}|<|\tilde{Q}_d|$, two types of extreme solutions exist whose mass
is $\tilde{M}_{ex}^{(1-)}$ and $\tilde{M}_{ex}^{(2-)}$.
For $\tilde{M}<\tilde{M}_{ex}^{(1-)}$, the solution has  a cosmological horizon and represents the spacetime with a globally naked singularity.
For $\tilde{M}=\tilde{M}_{ex}^{(1-)}$, the solution has  a degenerate black hole horizon and a cosmological horizon and represents the extreme black hole spacetime. 
For $\tilde{M}_{ex}^{(1-)}<\tilde{M}<\tilde{M}_{ex}^{(2-)}$,  the solution has three horizons, which are inner,  black hole and  comsological horizons.
The solution represents the black hole spacetime.
For $\tilde{M}=\tilde{M}_{ex}^{(2-)}$, the solution has an inner horizon and a degenerate horizon and represents the spacetime with a globally naked singularity.
For $\tilde{M}>\tilde{M}_{ex}^{(2-)}$, the solution has a cosmological horizon and represents the spacetime with a globally naked singularity.

When $|\tilde{Q}|=|\tilde{Q}_{d}|$, there is a solution with a doubly degenerate horizon.
For $\tilde{M}\ne \tilde{M}_{d}$, the solution has a cosmological horizon and represents the spacetime with a globally naked singularity.
For $\tilde{M}=\tilde{M}_{d}$, the solution has a doubly degenerate horizon and represents the spacetime with a globally naked singularity.

When $|\tilde{Q}|>|\tilde{Q}_{d}|$, no extreme solution exists.
All the solutions have a cosmological horizon only and represent the spacetime with a globally naked singularity.

\subsubsection{$k=0$  case}

Since there is no horizon for the solution in the plus branch, all the solutions represent the spacetime with a globally naked singularity.

The $\tilde{M}$-$r$ diagrams of the minus branch solutions are shown in 
Figs.~\ref{fig:M-r-6-charged}(f) ($n=6$)
and \ref{fig:M-r-5-charged}(f) ($n=5$).
The function $U(r_{ex})$ is negative
semi-definite so that there is no extreme solution. 
The solution has
a cosmological horizon and represents the spacetime with a globally naked singularity
for any mass parameter.

\subsubsection{$k=-1$  case}

The $\tilde{M}$-$r$ diagrams  are shown in
Figs.~\ref{fig:M-r-6-charged}(g) ($n=6$, minus branch), 
\ref{fig:M-r-6-charged}(h) ($n=6$, plus branch),
\ref{fig:M-r-5-charged}(g) ($n=5$, minus branch), and
\ref{fig:M-r-5-charged}(h) ($n=5$, plus branch).
The properties of the
$\tilde{M}$-$r_h$ relations are the same as those in the $\Lambda=0$ and $k=-1$ case
qualitatively. In the minus branch, however, the structure of  the infinity is
different (See the conformal diagrams).

In the case of the plus branch with $\Lambda=0$ and $k=-1$, 
$\tilde{M}_B$ is always less than
$\tilde{M}_{ex}^{(1)}$ for $|\tilde{Q}|<|\tilde{Q}_{d}|$ and $n\geq 6$, as is seen in Sec.~\ref{GB0}.
This is also true  in the present case.
By the $\tilde{M}$-$r_b$ relation (\ref{branch-sing}), $r_b$ is a monotonically decreasing function of $\tilde{M}$, and the horizon radius of the  plus-branch solution is
restricted as $r_h<\sqrt{2\tilde{\alpha}}$. By these facts the mass of the  solution which has a horizon should be greater than $\tilde{M}_B$. 
As a result, $\tilde{M}_B<\tilde{M}_{ex}^{(1+)}$ is obtained. 
This also holds in the $\Lambda < 0$ case.

Although it is difficult to show $|\tilde{Q}_{d}|<|\tilde{Q}_{B,0}|$
analytically,  we can show $|\tilde{Q}_{d}|<|\tilde{Q}_{B,0}|$  by comparing Eq.~(\ref{Q-tri}) with Eq.~(\ref{QB}) in the $\tilde{\alpha}\to 0$, $\tilde{\alpha}\to \infty$, and $n\to \infty$ limits. This supports the argument that this relation should hold for any $\tilde{\alpha}$ and $n$. Actually the numerical calculation shows $|\tilde{Q}_{d}|<|\tilde{Q}_{B,0}|$.
This implies that the negative mass black hole solution always exists for $|\tilde{Q}|\leq |\tilde{Q}_{d}|$.

\subsection{$\Lambda<0$}

\subsubsection{$k=1$  case}

The solution in the plus branch has no horizon and represents the spacetime with a  globally naked singularity. 

The $\tilde{M}$-$r$ diagrams of the minus-branch solutions are shown in 
Figs.~\ref{fig:M-r-6-charged}(i) ($n=6$)
and \ref{fig:M-r-5-charged}(i) ($n=5$).
Since the function $U(r_{ex})$ monotonically
increases from zero to infinity,
there is an extreme solution for any $\tilde{Q}$. 
The extreme solution has the positive mass $\tilde{M}^{(-)}_{ex}>0$ in the $n\geq 6$ case and $\tilde{M}^{(-)}_{ex}>\tilde{\alpha}$ in the $n=5$ case. 
For $\tilde{M}<\tilde{M}^{(-)}_{ex}$, the solution has no horizon and represents the spacetime with a  globally naked singularity. 
For $\tilde{M}=\tilde{M}^{(-)}_{ex}$, the solution has a degenerate black hole horizon and represents the extreme black hole spacetime. 
For $\tilde{M}>\tilde{M}^{(-)}_{ex}$, the solution has  an inner and a black hole horizon and represents the black hole spacetime.

\subsubsection{$k=0$  case}

The solution in the plus branch has no horizon and represents the spacetime with a  globally naked singularity. 

The $\tilde{M}$-$r$ diagrams of the minus-branch solutions are shown in 
Figs.~\ref{fig:M-r-6-charged}(j) ($n=6$)
and \ref{fig:M-r-5-charged}(j) ($n=5$).
The qualitative properties of the $\tilde{M}$-$r$ diagram are the same as those in the $k=1$ case except that the mass of the extreme solution is limited as $\tilde{M}_{ex}>0$ in the $n=5$ case.

\subsubsection{$k=-1$  case}

In $n\geq 6$ the function $U(r_{ex})$ has two extrema.
However,  $\tilde{Q}_d^2$ becomes negative for  the minus sign in Eq.~(\ref{BB}),
and eventually there is one extreme solution with a doubly degenerate horizon.
The value of the  charge $\tilde{Q}_d$ of the extreme solution is
determined by Eq.~(\ref{Q-tri}), where the plus sign is chosen in Eq.~(\ref{BB}).
When $|\tilde{Q}|<|\tilde{Q}_d|$, Eq.~(\ref{ex-pot}) has three roots, and there are three extreme solutions. One of them has
a degenerate horizon whose radius is large than $\sqrt{2\tilde{\alpha}}$ and belongs to the minus branch. The other belong to the plus branch.
When $|\tilde{Q}|=|\tilde{Q}_d|$, the degenerate horizons in the plus branch
coincide and become a doubly degenerate horizon. There is an extreme solution
in each branch.
When $|\tilde{Q}|>|\tilde{Q}_d|$, there is one extreme solution only in the minus branch. 
In the $n=5$ case, there is only one extreme solution for fixed $\tilde{Q}$ by Eq.~(\ref{ex-pot}). 
The radius of the degenerate horizon is $r_{ex}>\sqrt{2\tilde{\alpha}}$, and the extreme solution belongs to the minus branch.

First let us consider the minus branch.
The $\tilde{M}$-$r$ diagrams are shown in Figs.~\ref{fig:M-r-6-charged}(k) ($n=6$)
and \ref{fig:M-r-5-charged}(k) ($n=5$).
For any value of $\tilde{Q}$, there is an extreme solution with $\tilde{M}=\tilde{M}_{ex}^{(-)}$.
For $\tilde{M}<\tilde{M}_{ex}^{(-)}$, the solution has no horizon and represents
the spacetime with a globally naked singularity.
For $\tilde{M}=\tilde{M}_{ex}^{(-)}$, the solution has a degenerate horizon and represents the extreme black hole spacetime.
For $\tilde{M}_{ex}^{(-)}<\tilde{M}<\tilde{M}_{B}$, the solution has an inner and a black hole horizon and represents the   black hole spacetime.
For $\tilde{M}\geq\tilde{M}_{B}$, the solution has a black hole horizon and represents the  black hole spacetime. 

We plot the $\tilde{M}$-$r_{ex}$ curve on the $\tilde{M}$-$r$ diagrams.
For a small charge value, the extreme black hole solution has negative mass (the dot E in Figs.~\ref{fig:M-r-6-charged}(k) and \ref{fig:M-r-5-charged}(k)).
As the charge becomes large, the mass of the extreme black hole solution increases
and becomes zero. We denote the charge of the zero mass extreme solution $\tilde{Q}_{ex,0}$. Then 
when $|\tilde{Q}|<|\tilde{Q}_{ex,0}|$, the mass of the extreme black hole solution is negative, and the solution with $\tilde{M}_{ex}^{(-)}<\tilde{M}<0$ represents the   black hole spacetime with the negative mass.
When $|\tilde{Q}|=|\tilde{Q}_{ex,0}|$, the extreme solution represents the  zero mass black hole spacetime.
When $|\tilde{Q}|>|\tilde{Q}_{ex,0}|$, all the black hole solutions have the positive mass.

We have seen that several extreme solutions appear in Gauss-Bonnet gravity.
However, all of them do not have zero mass
except for the one in the case of $\Lambda<0$, $k=-1$ in the plus branch.
Let us examine this situation and derive the explicit form of $\tilde{Q}_{ex,0}$.
By the condition $\tilde{M}_{ex}=0$ (we assume $r_{ex}\ne 0$), 
Eq.~(\ref{M-extreme}) gives 
\begin{equation}
\label{QE-eq}
\frac{(n-2)}{(n-3)\ell^2}r_{ex}^4
+kr_{ex}^2
+\frac{(n-4)\tilde{\alpha}k^2}{(n-3)}
=0.
\end{equation}  
For $\Lambda=0$, 
\begin{equation}
r_{ex}^2 = -\frac{(n-4)\tilde{\alpha}k}{n-3},
\end{equation}  
which implies that only the $k=-1$ case can have a positive real root.
However, then the $\tilde{Q}_{ex,0}$ is estimated by Eq.~(\ref{Q-extreme}) as
\begin{equation}
\tilde{Q}_{ex,0}^2 = -\frac{2r_{ex}^{2(n-2)}}{(n-4)^2\tilde{\alpha}}<0,
\end{equation}  
which gives a contradiction so that $\tilde{Q}_{ex,0}$ is not defined in the $\Lambda=0$
case.

For $\Lambda \ne 0$, the roots of Eq.~(\ref{QE-eq}) are
\begin{equation}
\label{QE-3}
r_{ex}^2 = -\frac{(n-3)\ell^2}{2(n-2)}
\biggl[k\pm |k|\sqrt{1-\frac{4(n-2)(n-4)\tilde{\alpha}}{(n-3)^2\ell^2}} 
\biggr].
\end{equation}  
The value inside of the square root is guaranteed to be positive definite by the condition
(\ref{vacuum}).
For $k=0$ there is no positive root. Substituting Eq.~(\ref{QE-3}) into the
Eq.~(\ref{Q-extreme}), 
\begin{equation}
\label{QE-2}
\tilde{Q}_{ex,0}^2 =
-\frac{2r_{ex}^{2(n-4)}(kr_{ex}^2+2\tilde{\alpha}k^2)}{(n-2)(n-3)}.
\end{equation}  
When $k=1$ the r.h.s is negative definite. Hence only in the $k=-1$ case can
$\tilde{Q}_{ex,0}$ be defined. Moreover for $k=-1$, Eq.~(\ref{QE-2}) gives a condition
\begin{equation}
r_{ex}\bigl|_{\tilde{M}_{ex}=0} >\sqrt{2\tilde{\alpha}},
\end{equation}  
which implies that
the extreme solution with zero mass appears only in the minus branch.
Substituting Eq.~(\ref{QE-3}) into this condition in the $\Lambda>0$ case, one
easily finds
$4\tilde{\alpha} / \ell^2\geq 1$, which contradicts the condition (\ref{vacuum}).
In the $\Lambda<0$ case, taking the lower (minus sign) solution of Eq.~(\ref{QE-3}),
one finds a similar contradiction. As a result, only in the $\Lambda<0$ and
$k=-1$ case can we define  $\tilde{Q}_{ex,0}$ in the minus branch by taking the
upper solution of Eq.~(\ref{QE-3}). Its value is
\begin{equation}
\tilde{Q}_{ex,0}^2 =
\frac{2r_{ex}^{2(n-4)}(r_{ex}^2-2\tilde{\alpha})}{(n-2)(n-3)},
\end{equation}  
where
\begin{equation}
r_{ex}^2 = \frac{(n-3)\ell^2}{2(n-2)}
\biggl[1+ \sqrt{1-\frac{4(n-2)(n-4)\tilde{\alpha}}{(n-3)^2\ell^2}} 
\biggr].
\end{equation}  

Next let us move onto the plus branch.
The $\tilde{M}$-$r$ diagrams of the plus-branch solutions are shown in 
Figs.~\ref{fig:M-r-6-charged}(l) ($n=6$) and \ref{fig:M-r-5-charged}(l) ($n=5$). The configuration of the $\tilde{M}$-$r$ diagrams are qualitatively  the same as those in the $\Lambda=0$, $k=-1$  case, respectively.

In the $\Lambda \geq 0$ and $k=-1$ case, $|\tilde{Q}_d|<|\tilde{Q}_{B,0}|$ holds for any case. In the $\Lambda < 0$ case, however, $|\tilde{Q}_d|$ can be larger than $|\tilde{Q}_{B,0}|$. This can be seen easily as follows. In the $\alpha \to 0$ limit, 
\begin{eqnarray}
&& \tilde{Q}_{B,0}^2 \sim \frac{2^{n-3}}{n-3}\tilde{\alpha}^{n-3},
\\
&& \tilde{Q}_d^2 \sim 
\frac{2(n-5)}{(n-3)^3}\biggl[\frac{(n-4)(n-5)}{(n-3)^2}\biggr]^{n-4}
\tilde{\alpha}^{n-3}.
\end{eqnarray}  
So in this limit, $|\tilde{Q}_d|<|\tilde{Q}_{B,0}|$. On the other hand, in the 
$4\tilde{\alpha} /\ell^2 \to 1$ limit, $|\tilde{Q}_{B,0}| \sim 0$ and 
$|\tilde{Q}_d| >0$. Hence $|\tilde{Q}_d|>|\tilde{Q}_{B,0}|$.
Therefore there is a certain value of $\tilde{\alpha}$ where the $|\tilde{Q}_{B,0}|$ and $|\tilde{Q}_d|$
are equal. 
$|\tilde{Q}_{ex,0}|$ also vanishes in the $4\tilde{\alpha} /\ell^2 \to 1$ limit. 
Hence
for a small $\tilde{\alpha}$, there are two extreme solutions in the plus branch for small
$|\tilde{Q}|<|\tilde{Q}_d|$. Simultaneously the negative mass black hole solution exists.
As the charge becomes large, extreme solutions disappear while the negative mass black hole solution remains. 
In the large $\tilde{\alpha}$ limit, however, the negative mass black hole solutions disappear, although the extreme solutions exist. These behaviors in the space of the solutions
are interesting in the context of our discussion of the evolution of the black hole.

\section{Einstein-Gauss-Bonnet-Maxwell-$\Lambda$ system:
$4\tilde{\alpha}/\ell^2=1$ case}
\label{EGBML2}

The system with $4\tilde{\alpha}/\ell^2=1$ is a special case,
where the vacuum states of the plus and the minus branches coincide.
Since we have assumed that $\alpha$ is positive,
the cosmological constant is negative by definition.
The metric function $f$ becomes
\begin{equation}
f=k+\frac{r^2}{2\tilde{\alpha}}
\Biggl\{1\mp\sqrt{4\tilde{\alpha}
\biggl[\frac{\tilde{M}}{r^{n-1}}
-\frac{(n-3)\tilde{Q}^2}{2r^{2(n-2)}}\biggr]} \Biggr\}.
\end{equation}
When the radius is large the spacetime approaches adS spacetime  for $r\to \infty$ with the
effective curvature radius $\ell_{\rm eff}:=\sqrt{2\tilde{\alpha}}$ for $k=1$.

\begin{widetext}

\begin{figure}
\includegraphics[width=.95\linewidth]{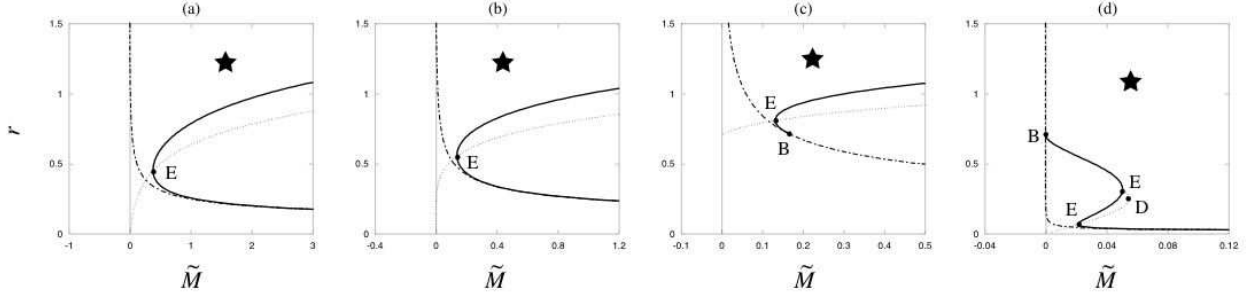}
\caption{
The $\tilde{M}$-$r$ diagrams of the static
solutions in the six-dimensional
Einstein-Gauss-Bonnet-Maxwell-$\Lambda$ system
with $4\tilde{\alpha}/\ell^2=1$. 
From the left-hand side,
the diagrams are the $k=1$ (the minus branch), $k=0$ (the minus branch), $k=-1$ (the minus branch), and  $k=-1$ (the plus branch) cases.
We show
the $\tilde{M}$-$r_h$ relations (thick solid curves) 
and the $\tilde{M}$-$r_b$ relations (dot-dashed curves).
We set $\tilde{\alpha}=0.25$ and $\tilde{Q}=0.1$ for (a) and (b),
$\tilde{Q}=0.2$ for (c), and  $\tilde{Q}=10^{-3}$ for (d).
We choose the value of the charge as $|\tilde{Q}|<|\tilde{Q}_d|$ for (d).
Otherwise, the two extreme solutions in the plus branch coincide, and the
$\tilde{M}$-$r_h$ relation becomes single valued. The dotted curve represents a sequence of the degenerate horizon ($\tilde{M}$-$r_{ex}$ relation)by varying $\tilde{Q}$.
Below the $\tilde{M}$-$r_b$ line or the negative mass regions, 
there are
no solutions. 
See Figs.~\ref{fig:E_m-rh} and \ref{fig:M-r-6-charged}  for the
meanings of the dots and the stars.
}
\label{fig:M-r-spe}
\end{figure}

\end{widetext}


In the charged case, a branch singularity  locates at
\begin{equation}
r=r_b:=
\biggl[\frac{(n-3)\tilde{Q}^2}{2\tilde{M}}\biggr]^{\frac{1}{n-3}}.
\end{equation}
Hence  the mass must be positive in this special case. 
There is no negative mass black hole solution.
Furthermore the $r_b$ is independent of the Gauss-Bonnet coefficient $\tilde{\alpha}$.
The Kretschmann invariant behaves as
\begin{equation}
{\cal I}=O((r-r_b)^{-3})
\end{equation}
around the branch singularity.
Since $f$ behaves 
\begin{equation}
f\approx k+\frac{r_b^2}{2\tilde{\alpha}}
\mp \frac{(n-3)|\tilde Q|}{\sqrt{2\tilde{\alpha}}r_b^{n-4}}
\sqrt{\frac{r-r_b}{r_b}}
\end{equation}
near the branch singularity, the
branch singularities are timelike (Fig.~\ref{penrose-center}(a)) for the solutions with $k=0,~1$. For the
solutions with
$k=-1$, the singularity is timelike for $r_b>\sqrt{2\tilde{\alpha}}$ and spacelike 
(Fig.~\ref{penrose-center}(d)) for $r_b<\sqrt{2\tilde{\alpha}}$.
In the special case of $r_b=\sqrt{2\tilde{\alpha}}$, the singularity is spacelike in the minus branch and  timelike in the plus branch.


Since the radius of a horizon  is restricted to
$r_h^2<-2\tilde{\alpha}k$ ($r_h^2>-2\tilde{\alpha}k$) for the plus (minus) branch,
there are no horizons
for solutions with $k=0,~1$ in the plus branch.
The $\tilde{M}$-$r_h$ relation (\ref{M-horizon}) becomes 
\begin{equation}
\tilde{M}
=r_h^{n-1}\biggl[
\frac{1}{4\tilde{\alpha}}
\biggl(1+\frac{2k\tilde{\alpha}}{r_h^2}\biggr)^2
+\frac{(n-3)\tilde{Q}^2}{2r_h^{2(n-2)}}
\biggr].
\end{equation}

As in the $4\tilde{\alpha}/\ell^2\ne 1$ case,
a vertical point of the $\tilde{M}$-$r_h$ curve corresponds to a degenerate horizon.
From the condition 
$f(r_{ex})=df/dr|_{r=r_{ex}}=0$, the relations between the mass, the charge, and the radius of the degenerate horizon of the extreme solution are
\begin{equation}
\tilde{M} =\tilde{M}_{ex}
=\frac{r_{ex}^{n-1}}{2\tilde{\alpha}(n-3)}
\biggl(1+\frac{2k\tilde{\alpha}}{r_{ex}^2}\biggr)
\biggl[n-2+\frac{2(n-4)k\tilde{\alpha}}{r_{ex}^2}
\biggr],
\end{equation}
\begin{equation}
\label{special_Q_ex}
\tilde{Q^2} =\tilde{Q}_{ex}^2
=\frac{r_{ex}^{2(n-2)}}{2\tilde{\alpha}(n-3)^2}
\biggl(1+\frac{2k\tilde{\alpha}}{r_{ex}^2}\biggr)
\biggl[n-1+\frac{2(n-5)k\tilde{\alpha}}{r_{ex}^2}
\biggr].
\end{equation}
There is one extreme solution for any value of $\tilde{Q}$ when $k=0,~1$ since the
r.h.s. of Eq.~(\ref{special_Q_ex}) monotonically increases. 
When $k=-1$ and $n=5$, there is also one extreme solution for any value of $\tilde{Q}$. Since the radius of the degenerate horizon is restricted to $r_{ex}>\sqrt{2\tilde{\alpha}}$ by Eq.~(\ref{special_Q_ex}) for the
$k=-1$ and $n=5$ case, the extreme solution belongs to the minus branch.

When $k=-1$ and $n\geq 6$,
the number of extreme solutions depends on $\tilde{Q}$.
There is always one extreme solution with $\tilde{M}=\tilde{M}_{ex}^{(-)}$ in
the minus branch. 
For $|\tilde{Q}|=|\tilde{Q}_{d}|$, where
$|\tilde{Q}_{d}|$ is obtained by substituting
$\ell^2=4\tilde{\alpha}$ into Eq.~(\ref{Q-tri}),
there is an extreme solution with a
doubly degenerate horizon in the plus branch.
$\tilde{M}_d$ and $r_{d}$
are also obtained from Eqs.~(\ref{r-tri}), (\ref{M-tri}),  and (\ref{BB}).
For $|\tilde{Q}|< (>) |\tilde{Q}_{d}|$ 
there are two (no) extreme solutions with 
 $\tilde{M}_{ex}^{(1+)}$ and $\tilde{M}_{ex}^{(2+)}$,  where $\tilde{M}_{ex}^{(1+)}<\tilde{M}_{ex}^{(2+)}$, respectively,
in the plus branch.


By the above analysis, we can draw the $\tilde{M}$-$r$ diagram.
The solution in the plus branch has no horizon in the $k=1,~0$ cases
and represents the spacetime with a globally naked singularity.
The $\tilde{M}$-$r$ diagrams  of the minus-branch solution
in the $k=1,~0$ cases are shown in Figs.~\ref{fig:M-r-spe}(a) and \ref{fig:M-r-spe}(b),
respectively. 
The solution only exists for the positive mass parameter.
Besides this fact, the
properties are the same as those in the $4\tilde{\alpha}\ne \ell^2$ case
(see Figs.~\ref{fig:M-r-6-charged}(i) and \ref{fig:M-r-6-charged}(j)).
There is an extreme solution for any $\tilde{Q}$.
For $0<\tilde{M}<\tilde{M}^{(-)}_{ex}$, the solution has no horizon
and represents the spacetime with a globally naked singularity.
For $\tilde{M}=\tilde{M}^{(-)}_{ex}$, the solution has a degenerate black hole horizon
and represents the extreme black hole spacetime.
For $\tilde{M}>\tilde{M}^{(-)}_{ex}$, the solution has an inner and a black hole horizon and represents the black hole spacetime.

In the $k=-1$ case, the $\tilde{M}$-$r$ diagram of the minus-branch solution
is shown in Fig.~\ref{fig:M-r-spe}(c).
There is an extreme solution for any $\tilde{Q}$.
For $0<\tilde{M}<\tilde{M}^{(-)}_{ex}$, the solution has no horizon
and represents the spacetime with a globally naked singularity.
For $\tilde{M}=\tilde{M}^{(-)}_{ex}$, the solution has a degenerate black hole horizon
and represents the extreme black hole spacetime.
For $\tilde{M}^{(-)}_{ex}<\tilde{M}<\tilde{M}_{B}$, the solution has inner and  black hole horizons
and represents the black hole spacetime.
For $\tilde{M}_{B}\leq\tilde{M}$, the solution has a black hole horizon
and represents the black hole spacetime.

The $\tilde{M}$-$r$ diagram of the plus-branch solution
is shown in Fig.~\ref{fig:M-r-spe}(d).
The configuration of the $\tilde{M}$-$r_h$ curve is almost the same as that in  the 
$4\tilde{\alpha}/\ell^2\ne 1$ case (see Fig.~\ref{fig:M-r-6-charged}(l)) except that the mass parameter is always positive, and eventually the mass of the branch point is positive.

The spacetime structures of these solutions are summarized in
Tables~\ref{GB-penrose-6-charged_2} and  \ref{GB-penrose-spe-charged}.

\section{Conclusions and discussion}
\label{Conclusion}

We have studied spacetime structures of the static solutions in
the $n$-dimensional Einstein-Gauss-Bonnet-Maxwell-$\Lambda$ system, where the
cosmological constant is either positive, zero, or negative.
We assume that the Gauss-Bonnet coefficient $\alpha$ is non-negative. This assumption is consistent with the notion that the action is derived from superstring/M-theory in the low-energy limit.
The solutions  have the $(n-2)$-dimensional Euclidean sub-manifold whose  curvature  is $k=1,~0$, or $-1$. 
We assume $4{\tilde \alpha}/\ell^2\leq 1$ in order to define the relevant vacuum state.
The structures of the center, horizons,  infinity, and the singular point depend on the parameters of the system and the branches complicatedly so that a variety of global structures for the solution is found. 
In our analysis, the $\tilde{M}$-$r$ diagram provides an  easily understood visual clatification.

In Gauss-Bonnet gravity, the solutions are classified into plus and minus branches. In the $\tilde{\alpha}\to 0$ limit, 
the solution in the minus branch recovers the one in general relativity, while there is no solution in the plus branch.
The structure of  the infinity is not determined by the cosmological constant alone but also by the effective curvature radius 
$\ell_{\rm eff}^2$. In the minus branch the ordinary correspondence between the signs of the cosmological constant $\Lambda$ and the asymptotic structures is obtained.
In the plus branch, however, the sign
of $\ell_{\rm eff}^2$ is always positive independent of the cosmological constant,
so that all the solutions have the same  asymptotic structure as those in general relativity with a negative cosmological constant. 

In general relativity, when one adds a charge to the Schwarzschild black hole,
the inner horizon appears, and the singularity changes from spacelike to timelike.
Also in Gauss-Bonnet gravity, 
the charge affects greatly the central region of the spacetime.
In the charged solutions a singularity appears at the finite radius $r=r_b>0$, which is called a branch singularity. Although in the neutral case the branch singularity appears only for the negative mass parameter, in the charged case it appears for any mass parameter. It becomes timelike or spacelike depending on the parameters.
Furthermore, the Kretschmann invariant behaves as 
$O((r-r_b)^{-3})$ around the branch singularity. This is much milder than
divergent behavior of the central singularity in general relativity $O(r^{-4(n-2)})$.
This may imply that the string effects make the singular behavior milder.

There are three types of horizon: inner,  black hole, and  cosmological horizons.
In the $k=1,~0$ cases the plus-branch solutions do not have any horizon.
In the $k=-1$ case, the radius of the horizon is restricted as $r_h<\sqrt{2\tilde{\alpha}}$ ($r_h>\sqrt{2\tilde{\alpha}}$) in the plus (minus)
branch. 
The point with $r_h=\sqrt{2\tilde{\alpha}}$ on the $\tilde{M}$-$r_h$ relation
is the branch point whose mass is $\tilde{M}_B$. In the neutral case $\tilde{M}_B$ is always negative, so that there is a black hole solution with negative mass. However, in the charged case $\tilde{M}_B$ becomes zero when
$|\tilde{Q}|=|\tilde{Q}_{B,0}|$, and  $\tilde{M}_B>0$ for $|\tilde{Q}|>|\tilde{Q}_{B,0}|$.

Among the solutions in our analysis, the black hole solutions would be the most important. Although the $\tilde{M}$-$r_h$ relations in the $k=1,~0$ cases have qualitatively
similar configurations to those in general relativity, the solution in the $k=-1$ case
is quite different.
The charged black hole solution in general relativity always  has an inner horizon,
and the singularity is timelike. 
For the $k=1$ and $0$ cases in Gauss-Bonnet gravity, the singularity of the
black hole solution is timelike, too.
Most of
the black hole solutions with $k=-1$, however, have no inner horizon so that
the branch singularity of these solutions becomes spacelike.
When $|\tilde{Q}|<|\tilde{Q}_{B,0}|$, there are black hole solutions with zero and negative mass in the plus branch for $k=-1$ regardless of the sign of the
cosmological constant. Although there is maximum mass for the black hole solutions  in the plus branch for $k=-1$ in the neutral case, no such maximum exists in the charged case, and any positive mass solution is black hole for  $|\tilde{Q}|<|\tilde{Q}_{B,0}|$.
A solution has three horizons at most. In general relativity, the RN-dS black hole solution has three horizons which are the inner, the black hole,  and the cosmological
horizons. In Gauss-Bonnet gravity, the solutions in the plus branch with $k=-1$ and  $n\geq6$ have three horizons for each cosmological constant when $|\tilde{Q}|<|\tilde{Q}_d|$ and 
$\tilde{M}^{(1+)}_{ex}<\tilde{M}<\tilde{M}^{(2+)}_{ex}$. The horizons of these solutions are the ``inner" black hole, the inner, and the ``outer" black hole horizons.
We add the words ``inner" and ``outer" to distinguish each black hole horizon.

We investigate the system with the special parameter $4{\tilde \alpha}/\ell^2=1$
separately where the vacuum states in the plus and minus branches coincide.
In this case only is the positive mass solution  allowed; otherwise the metric function
takes a complex value.

Considering the Gauss-Bonnet effects on the CCH, we find firstly that the divergent behavior around the singularity becomes milder than that in general relativity. It is, however,  difficult to eliminate the singularity completely. 
Secondly, most of the black hole solutions in the $k=-1$ case have no inner horizon, and the singularity is spacelike. 
However, all the black hole solutions in $k=1$ and $0$ cases still have the timelike singularity. 
Furthermore since the structure of  the infinity of the black hole solution with $k=-1$ is 
adS-like, the spacetime is not globally hyperbolic. As a result, the Gauss-Bonnet terms do not work quite well from the cosmic censorship spirit.

Let us consider the evolution of the black hole solutions. In the RN-dS black hole spacetime in general relativity, the spacetime  seems to evolve to the one with globally naked 
singularity by throwing some mass into the black hole such that the mass of the black hole becomes larger than that of the extreme black hole solution $\tilde{M}_{ex}^{(2)}$. However, it will not happen as discussed in Ref.~\cite{Nakao}.
In Gauss-Bonnet gravity, the situation is quite different. In the non-extreme neutral black hole spacetime in the plus branch with $k=-1$,  there are inner and black hole horizons without a cosmological horizon~\cite{TM}. When one put the mass into the black hole, the radius of black hole decreases~\cite{note}. If more matter is added,
the black hole may evolve to the extreme black hole with finite processes and then to the spacetime with a globally naked singularity. This is a classical transition from the black hole spacetime to the one with a globally naked singularity.
Next let us consider the charged case where the black hole spacetime has the ``inner" black hole, and
the inner and the ``outer" black hole horizons. This solution exists in the plus branch for $n\geq 6$. Adding the mass to the black hole,
the black hole approaches the extreme solution, and then it transits to the black hole spacetime with a smaller horizon whose mass is $\tilde{M}>\tilde{M}_{ex}^{(2+)}$.
This is a classical transition from one black hole spacetime to another black hole spacetime. 
On the other hand, another process, i.e., the Hawking evaporation process, may occur
in our black hole spacetime. The black hole in this plus-branch solution may evaporate and lose its mass. Then, what happens when the mass of the black hole becomes $\tilde{M}=\tilde{M}_{ex}^{(1+)}$, where the horizon inside of the ``outer" black hole horizon degenerates?
To obtain the precise scenario of  the evolutions of black hole spacetime through classical and quantum processes, we need detailed analysis.

There are some applications and extensions of the present investigation.
We find that the static solutions in Gauss-Bonnet gravity have much variety and interesting properties. One of the most important issues is stability of the solutions. There are some studies which support the dynamical stability of the exterior spacetime of the black hole solutions in Gauss-Bonnet gravity~\cite{Neupane,Dotti,Konoplya}. However, 
they do not cover all the cases, and there remain some uncleared analyses.
The problem of the stability of the inner horizons has also been raised. This is interesting from
the CCH point of view~\cite{mass-inf,kink}.
We find several solutions with degenerate horizons. They correspond to the solutions with product metric such as the Nariai and the  Bertotti-Robinson solutions. All the solutions of this type in the Einstein-Maxwell-$\Lambda$ system have been  recently classified in Ref.~\cite{Lemos}. It would be interesting to extend the analysis to Gauss-Bonnet gravity.  Another application is to the braneworld. The solution with a negative cosmological constant and/or in the plus branch has the adS-like structure of  the infinity. These solutions can be applied to the  bulk spacetime of the braneworld and
are currently under investigation.

\section*{Acknowledgements}

We would like to thank 
Kei-ichi Maeda 
and 
Umpei Miyamoto 
for their discussion. 
This work was partially supported by a Grant for The 21st Century COE Program (Holistic Research and Education Center for Physics Self-Organization Systems) at Waseda University.


\newpage


\begin{figure}
\includegraphics[width=.95\linewidth]{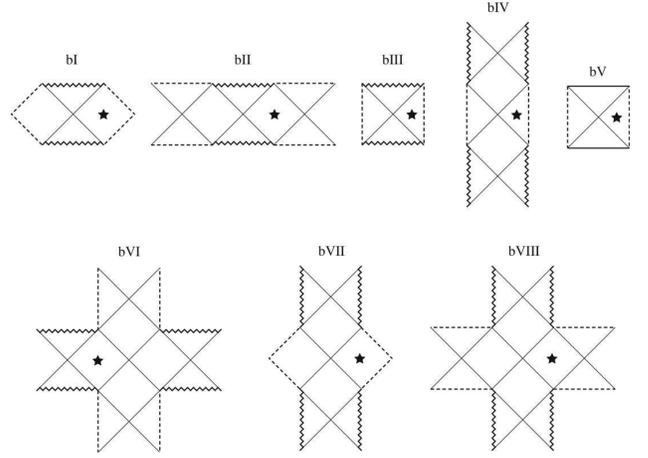}
\caption{
The conformal diagrams of black hole solutions.
bI, bII, bIII, bIV, bVII, and bVIII have the same structure as those of 
the Schwarzschild, 
Schwarzschild-dS, Schwarzschild-adS, RN-adS, RN,
and RN-dS spacetimes, respectively.
}
\label{penrose_BH}
\end{figure}

\begin{figure}[h]
\includegraphics[width=.62\linewidth]{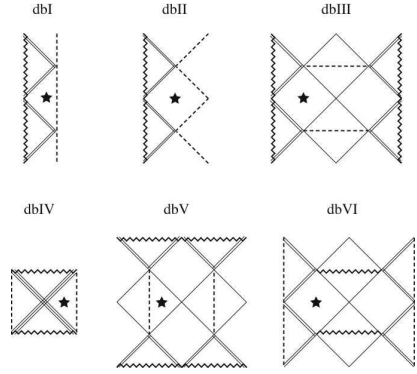}
\caption{
The conformal diagrams of black hole solutions
with degenerate horizons. 
The degenerate horizon is drawn with a double or triple line according to its
degeneracy. By passing across the triple (double)
line, the trapped  region (does not) changes to a untrapped one.
dbI, dbII, and dbIII have the same structure as those of
the extreme RN-adS, extreme RN,
and extreme RN-dS spacetimes with degenerated horizons of the
inner and black hole horizons, respectively. dbIV is the case with triple degenerate horizons.
Although the horizontal lines in dbIII are the infinities, they show distinct infinities by a
single line depending on how
to approach (from above or from below) them.
}
\label{penrose_BH_degenerate}
\end{figure}

\begin{figure}[t]
\includegraphics[width=.80\linewidth]{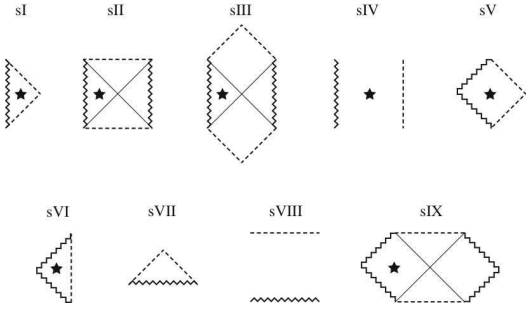}
\caption{
The conformal diagrams of the solutions with a globally naked singularity.
There are time-reversed diagrams for sVII and sVIII.
}
\label{penrose_singular}
\end{figure}

\begin{figure}[t]
\includegraphics[width=.75\linewidth]{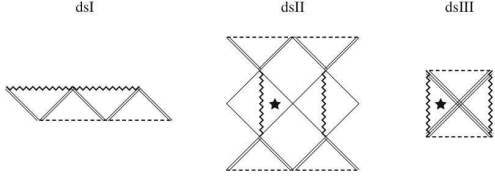}
\caption{
The conformal diagrams of the solutions with a globally naked singularity and
degenerate horizons.
The timelike singularities in dsII show distinct singularities by a single wavy line depending on
how to approach (from the right or left hand side) to them.
There is a time-reversed diagram for dsI.
}
\label{penrose_sing_degenerate}
\end{figure}

\newpage

 
\begin{widetext}

\begin{table}
\renewcommand{\arraystretch}{1.2}
\caption{
The classification of the spacetime structures of the static
solutions in the Einstein-Maxwell-$\Lambda$ system.
The numbers in the column ``type" imply the types of the conformal
diagrams in
Figs.~\ref{penrose_BH}--\ref{penrose_sing_degenerate}.
For the $\Lambda>0$ and $k=1$ case, see Table~\ref{E-penrose-charged_2}.
}
\label{E-penrose-charged}
\vspace{4mm}
\begin{tabular}{c|c|c|c|c|c|c}
\hline
\hline
\lw{~$k$~} &
\multicolumn{2}{c|}{$\Lambda=0$} & 
\multicolumn{2}{c|}{$\Lambda>0$} &
\multicolumn{2}{c}{$\Lambda<0$}
\\
\cline{2-7}
 & ~~~~~~~~~~$\tilde{M}$~~~~~~~~~~ & ~type~ &
   ~~~~~~~~~~$\tilde{M}$~~~~~~~~~~ & ~type~ &
   ~~~~~~~~~~$\tilde{M}$~~~~~~~~~~ & ~type~ 
\\
\hline \hline 
1 & $\tilde{M}<\tilde{M}_{ex}$ & sI & 
see Table~\ref{E-penrose-charged_2}&   &
$\tilde{M}<\tilde{M}_{ex}$ & sIV
\\
\cline{2-7}
   & $\tilde{M}=\tilde{M}_{ex}$ & dbI & 
 &    &
$\tilde{M}=\tilde{M}_{ex}$ & dbII
\\
\cline{2-3}\cline{6-7}
   & $\tilde{M}>\tilde{M}_{ex}$ & bVII & 
 &    &
$\tilde{M}>\tilde{M}_{ex}$ & bIV
\\
\noalign{\hrule height 1.0pt}
 0  & $\tilde{M}\leq 0$ & sI & 
any &  sII & 
$\tilde{M}<\tilde{M}_{ex}$ & sIV
\\
\cline{2-7}
  & $\tilde{M}>0$ & sIII & 
 &  & 
$\tilde{M}=\tilde{M}_{ex}$ & dbII
\\
\cline{2-3} \cline{6-7}
  &  &  & 
 &  & 
$\tilde{M}>\tilde{M}_{ex}$ & bIV
\\
\noalign{\hrule height 1.0pt}
 -1  & any &  sIII & 
 any &  sII  & 
$\tilde{M}<\tilde{M}_{ex}$ & sIV
\\
\cline{2-7}
   &  &  & 
 &  & 
$\tilde{M}=\tilde{M}_{ex}$ & dbII
\\
\cline{6-7}
   &  &  & 
 &  & 
$\tilde{M}>\tilde{M}_{ex}$ & bIV
\\
\hline
\hline
\end{tabular}
\end{table}



\begin{table}
\renewcommand{\arraystretch}{1.2}
\caption{
The classification of the spacetime structures of the static
solutions in the $\Lambda>0$ and $k=1$ case 
in the Einstein-Maxwell-$\Lambda$ system.
This table also show the classification of the spacetime structures of the static
solutions in the minus branch in the  Einstein-Gauss-Bonnet-Maxwell-$\Lambda$ system with $4\tilde{\alpha}/\ell^2\ne1$. In the latter case the mass of the extreme solutions are 
$\tilde{M}^{(i)}_{ex} \to \tilde{M}^{(i-)}_{ex}$, where $i=1,\:2$.
The numbers in the column ``type" imply the types of the conformal
diagrams in
Figs.~\ref{penrose_BH}--\ref{penrose_sing_degenerate}.
}
\label{E-penrose-charged_2}
\vspace{4mm}
\begin{tabular}{c|c|c|c|c|c}
\hline
\hline

\multicolumn{2}{c|}{$|\tilde{Q}|<|\tilde{Q}_d|$} & 
\multicolumn{2}{c|}{$|\tilde{Q}|=|\tilde{Q}_d|$} &
\multicolumn{2}{c}{$|\tilde{Q}|>|\tilde{Q}_d|$}
\\
\cline{1-6}
 ~~~~~~~~~~$\tilde{M}$~~~~~~~~~~ & ~type~ &
   ~~~~~~~~~~$\tilde{M}$~~~~~~~~~~ & ~type~ &
   ~~~~~~~~~~$\tilde{M}$~~~~~~~~~~ & ~type~ 
\\
\hline \hline 
 $\tilde{M}<\tilde{M}_{ex}^{(1)}$ & sII & 
$\tilde{M}<\tilde{M}_{d}$ & sII &
any & sII 
\\
\cline{1-6}
 $\tilde{M}=\tilde{M}_{ex}^{(1)}$ & dbIII & 
$\tilde{M}=\tilde{M}_{d}$ & dsIII &
 &
\\
\cline{1-4}
 $\tilde{M}_{ex}^{(1)}<\tilde{M}<\tilde{M}_{ex}^{(2)}$ & bVIII & 
$\tilde{M}>\tilde{M}_{d}$ & sII &
 & 
\\
\cline{1-4}
 $\tilde{M}=\tilde{M}_{ex}^{(2)}$  & dsII & 
 &  & & 
\\
\cline{1-2}
$\tilde{M}>\tilde{M}_{ex}^{(2)}$  & sII & 
 & & & 
\\
\hline
\hline
\end{tabular}
\end{table}



\begin{table}
\renewcommand{\arraystretch}{1.2}
\caption{ 
The classification of the spacetime structures of the static
solutions 
in the Einstein-Gauss-Bonnet-Maxwell-$\Lambda$ system with 
$4\tilde{\alpha}/\ell^2\ne 1$.
The
numbers in the column ``type" imply the types of the conformal diagrams
in Figs.~\ref{penrose_BH}-\ref{penrose_sing_degenerate}.
For the minus-branch solutions with $\Lambda>0$ and $k=1$,
and  the plus-branch solutions with $k=-1$,
see Tables~\ref{E-penrose-charged_2} and ~\ref{GB-penrose-6-charged_2}, respectively.
}
\label{GB-penrose-6-charged}
\vspace{4mm}
\begin{tabular}{c|c|c|c|c|c|c|c|c}
\hline
\hline
\lw{~$k$~} &
\multicolumn{2}{c|}{$\Lambda=0$, ($-$ branch)} & 
\multicolumn{2}{c|}{$\Lambda>0$, ($-$ branch)} &
\multicolumn{2}{c|}{$\Lambda<0$, ($-$ branch)} &
\multicolumn{2}{c}{$+$ branch} 
\\
\cline{2-9}
&  ~~~~~~~~~~$\tilde{M}$~~~~~~~~~~ & ~type~ 
&   ~~~~~~~~~~$\tilde{M}$~~~~~~~~~~ & ~type~ 
&   ~~~~~~~~~~$\tilde{M}$~~~~~~~~~~ & ~type~ 
& ~~~~~~~~~~$\tilde{M}$~~~~~~~~~~ & ~type~ 
\\
\hline \hline 
1 & $\tilde{M}<\tilde{M}_{ex}^{(-)}$ & sI 
& see Table~\ref{E-penrose-charged_2} & 
& $\tilde{M}<\tilde{M}_{ex}^{(-)}$ & sIV 
& any & sIV 
\\
\cline{2-7}
& $\tilde{M}=\tilde{M}_{ex}^{(-)}$ & dbI
&  &  
& $\tilde{M}=\tilde{M}_{ex}^{(-)}$ & dbII 
& &
\\
\cline{2-3}\cline{6-7}
& $\tilde{M}>\tilde{M}_{ex}^{(-)}$ & bVII
&  &  
& $\tilde{M}>\tilde{M}_{ex}^{(-)}$ & bIV 
&  &
\\
\noalign{\hrule height 1.0pt}
0  & $\tilde{M}\leq 0$ & sI
& any & sII 
& $\tilde{M}<\tilde{M}_{ex}^{(-)}$ & sIV 
& any & sIV
\\
\cline{2-9}
& $\tilde{M}>0$ & sIII
&  &
& $\tilde{M}=\tilde{M}_{ex}^{(-)}$ & dbII
&  &
\\
\cline{2-3}\cline{5-7}
&  &
&  &
& $\tilde{M}>\tilde{M}_{ex}^{(-)}$ & bIV 
&  &
\\
\noalign{\hrule height 1.0pt}
-1  & $\tilde{M}<\tilde{M}_{B}$ & sIII
& $\tilde{M}<\tilde{M}_{B}$ & sII 
& $\tilde{M}<\tilde{M}_{ex}^{(-)}$ & sIV 
& see Table~\ref{GB-penrose-6-charged_2} & 
\\
\cline{2-9}
& $\tilde{M}\geq \tilde{M}_{B}$ & sVII
& $\tilde{M}\geq\tilde{M}_{B}$ & sVIII
& $\tilde{M}=\tilde{M}_{ex}^{(-)}$ & dbII 
&  & 
\\
\cline{2-7}
&  & 
&  & 
& $\tilde{M}_{ex}^{(-)}<\tilde{M}<\tilde{M}_{B}$ & bIV 
&  & 
\\
\cline{2-7}
&  & 
&  & 
&  $\tilde{M}\geq\tilde{M}_{B}$ & bIII 
&  & 
\\
\hline
\hline
\end{tabular}
\vspace{12pt}
\end{table}



\begin{table}
\renewcommand{\arraystretch}{1.2}
\caption{ 
The classification of the spacetime structures of the static
solutions in the plus-branch with $k=-1$ in the  
Einstein-Gauss-Bonnet-Maxwell-$\Lambda$ system. 
The numbers in the column ``type" imply the types of the conformal
diagrams in
Figs.~\ref{penrose_BH}--\ref{penrose_sing_degenerate}.
}
\label{GB-penrose-6-charged_2}
\vspace{4mm}
\begin{tabular}{c|c|c|c|c|c|c|c}
\hline
\hline

\multicolumn{2}{c|}{$|\tilde{Q}|<|\tilde{Q}_d|$ $(n=6)$}  & 
\multicolumn{2}{c|}{$|\tilde{Q}|=|\tilde{Q}_d|$ $(n=6)$} &
\multicolumn{2}{c|}{$|\tilde{Q}|>|\tilde{Q}_d|$ $(n=6)$} &
\multicolumn{2}{c}{$n=5$}
\\
\cline{1-8}
 ~~~~~~~~~~$\tilde{M}$~~~~~~~~~~ & ~type~ &
   ~~~~~~~~~~$\tilde{M}$~~~~~~~~~~ & ~type~ &
   ~~~~~~~~~~$\tilde{M}$~~~~~~~~~~ & ~type~ &
   ~~~~~~~~~~$\tilde{M}$~~~~~~~~~~ & ~type~ 
\\
\hline \hline 
 $\tilde{M}\leq\tilde{M}_{B}$ & sIV & 
$\tilde{M}\leq\tilde{M}_{B}$ & sIV &
$\tilde{M}\leq\tilde{M}_{B}$ & sIV &
$\tilde{M}\leq\tilde{M}_{B}$ & sIV
\\
\cline{1-8}
$\tilde{M}_{B}<\tilde{M}<\tilde{M}_{ex}^{(1+)}$ & bIII & 
$\tilde{M}_{B}<\tilde{M}<\tilde{M}_{d}$ & bIII &
$\tilde{M}>\tilde{M}_{B}$ & bIII &
$\tilde{M}>\tilde{M}_{B}$ & bIII
\\
\cline{1-8}
$\tilde{M}=\tilde{M}_{ex}^{(1+)}$ & dbV & 
$\tilde{M}=\tilde{M}_{d}$ & dbIV &
 &  & 
 & 
\\
\cline{1-4}
$\tilde{M}_{ex}^{(1+)}<\tilde{M}<\tilde{M}_{ex}^{(2+)}$ & bVI & 
$\tilde{M}>\tilde{M}_{d}$ & bIII &
 &  & 
 & 
\\
\cline{1-4}
$\tilde{M}=\tilde{M}_{ex}^{(2+)}$ & dbVII & 
 &  & 
 &  & 
 & 
\\
\cline{1-2}
$\tilde{M}>\tilde{M}_{ex}^{(2+)}$ & bIII & 
 &  & 
 &  & 
 & 
\\
\hline
\hline
\end{tabular}
\end{table}



\begin{table}
\renewcommand{\arraystretch}{1.2}
\caption{ 
The classification of the spacetime structures of the static solutions in the 
Einstein-Gauss-Bonnet-Maxwell-$\Lambda$
system with $4\tilde{\alpha}/\ell^2=1$.  The
numbers in the column ``type" imply the types of the Penrose diagrams
drawn in Figs.~\ref{penrose_BH}-\ref{penrose_sing_degenerate}.
For the plus branch solution with $k=-1$, see Table~\ref{GB-penrose-6-charged_2},
where the range $\tilde{M}\leq\tilde{M}_{B}$ should be changed to
$0<\tilde{M}\leq\tilde{M}_{B}$.
}
\label{GB-penrose-spe-charged}
\vspace{4mm}
\begin{tabular}{c|c|c|c|c}
\hline
\hline
\lw{~$k$~} 
&\multicolumn{2}{c|}{$-$ branch}
&\multicolumn{2}{c}{$+$ branch}
\\
\cline{2-5}
&  ~~~~~~~~~~$\tilde{M}$~~~~~~~~~~ & ~type~ 
&  ~~~~~~~~~~$\tilde{M}$~~~~~~~~~~ & ~type~ 
\\
\hline \hline 
1,~0
& $0<\tilde{M}<\tilde{M}_{ex}^{(-)}$ & sIV
& $\tilde{M}>0$ & sIV 
\\
\cline{2-5}
& $\tilde{M}=\tilde{M}_{ex}^{(-)}$ & dbII
&  & 
\\
\cline{2-3}
& $\tilde{M}>\tilde{M}_{ex}^{(-)}$ & bIV
&  & 
\\
\noalign{\hrule height 1.0pt}
-1 
& $0<\tilde{M}<\tilde{M}_{ex}^{(-)}$ & sIV
& \multicolumn{2}{c}{see Table~\ref{GB-penrose-6-charged_2}} 
\\
\cline{2-3}
& $\tilde{M}=\tilde{M}_{ex}^{(-)}$ & dbII
&  \multicolumn{2}{c}{(The range $\tilde{M}\leq\tilde{M}_{B}$ is~~~~~~}
\\
\cline{2-3}
& $\tilde{M}_{ex}^{(-)}<\tilde{M}<\tilde{M}_B$ & bIV
& \multicolumn{2}{c}{ ~~~changed to $0<\tilde{M}\leq\tilde{M}_{B}$.)~~}
\\
\cline{2-3}
& $\tilde{M}\geq\tilde{M}_B$ & bIII
&  \multicolumn{2}{c}{}
\\
\hline
\hline
\end{tabular}
\end{table}

\end{widetext}


\begin{references}
\bibitem{note_1} 
   In the higher-dimensional cases, the $(n-2)$-dimensional
Einstein space is not necessarily sphere or its quotient space even
for positive curvature. Hence the solution represents not only the (at least locally) RN spacetime but also
many others. However, we refer to all such cases together as 
``RN spacetimes".
In the following sections we will also follow this usage.
\bibitem{note} 
   Since entropy of the black hole increases, the decrease of the area of the black hole horizons does not contradict the second law of black hole thermodynamics.
\end{references}
\end{document}